\documentclass[12pt] {article}

\input epsf

\begin{document}\setlength{\unitlength}{1mm}
\def\question#1{{{\marginpar{\small \sc #1}}}}
\newcommand{\QCD}{{ \rm QCD}^{\prime}}
\newcommand{\MSSM}{{ \rm MSSM}^{\prime}}
\newcommand{\eq}{\begin{equation}}
\newcommand{\en}{\end{equation}}
\newcommand{\bino}{\tilde{b}}
\newcommand{\tsquark}{\tilde{t}}
\newcommand{\gluino}{\tilde{g}}
\newcommand{\photino}{\tilde{\gamma}}
\newcommand{\wino}{\tilde{w}}
\newcommand{\mtilde}{\tilde{m}}
\newcommand{\higgsino}{\tilde{h}}
\newcommand{\gsi}{\,\raisebox{-0.13cm}{$\stackrel{\textstyle>}
{\textstyle\sim}$}\,}
\newcommand{\lsi}{\,\raisebox{-0.13cm}{$\stackrel{\textstyle<}
{\textstyle\sim}$}\,}
\rightline{CERN-TH/99-130}
\rightline{hep-ph/9905441}

\baselineskip=18pt
\vskip 0.6in
\begin{center}
{ \LARGE Mixing Effects in the  Finite-Temperature Effective Potential of the MSSM with a Light Stop }\\
\vspace*{0.6in}
{\large Marta Losada\footnote{On leave of absence
from the Universidad Antonio Nari\~{n}o, Santa Fe de
Bogot\'a, COLOMBIA.}} \\
\vspace{.1in}
{\it CERN Theory Division \\ CH-1211
Geneva 23, Switzerland}\\
\vspace{.2in}
\vspace{.1in}
\end{center}
\vspace*{0.05in}
\vskip  0.2in  

Abstract: We incorporate the effects of mixing arising from the trilinear
terms in the MSSM potential to the  effective three dimensional theory for
the MSSM at high temperature  in the
limit of large $m_{A}$. There are relevant one-loop effects that modify
the 3D parameters of the effective theory.
 We calculate the two-loop effective potential of
the 3D theory for the Higgs and the right handed stop to
analyse the possible phase transitions and  to determine
the precise region in the $m_{h}$-$m_{\tilde{t}_{2}}$ plane
for which the sphaleron constraint  for preservation
of the baryon asymmetry is satisfied. There is an upper bound
on the value of the mixing parameter coming from stop searches. We also compare with 
previous results obtained using 4D calculations of the effective potential for the regime of large $m_{Q}$. A two-stage phase transition  persists
for a small range of values of $m_{\tilde{t}_{2}}$ for  given values of the mixing
parameter and $\tan\beta$. This can further constrain the allowed region of
parameter space.  Electroweak baryogenesis requires a
value of $m_{\tilde{t}_{2}}\lsi 170$ and $m_{h} \lsi 105$  GeV for $m_{Q}=300$ GeV.

\vskip 0.2in
\noindent
CERN-TH/99-130\\
May 1999

\thispagestyle{empty}
\newpage
\addtocounter{page}{-1}
\newpage

\section{Introduction}

The  region of parameter space for which the electroweak baryogenesis scenario is viable in the context
of the Minimal Supersymmetric Standard Model (MSSM)
is currently being tested at existing accelerators.
Electroweak baryogenesis requires a strong first-order electroweak phase transition. This ensures that sphaleron 
transitions are switched off in the broken phase, permitting
the preservation of a baryon asymmetry \cite{kuzmin, shaposhnikov} (for reviews, see \cite{ckn}-\cite{riotto}). In the context of the Standard Model the only free parameter relevant to the phase transition is the Higgs mass. Non-perturbative results
of the phase transition  show that for the Standard Model there is no experimentally allowed value of the Higgs mass for which the phase transition is sufficiently
first order \cite{KLRS, Fodor}\footnote{Non-perturbative analyses also show good agreement
with perturbation theory for Higgs mass values $m_{h}\lsi 70$GeV \cite{KLRS}.}.  Non-perturbative studies  of the phase transition are necessary at finite temperature
 due to the existence of infrared divergences in gauge theories. However,
recent lattice simulations have shown that for the case of the MSSM the perturbative 
results are conservative in the constraints they impose on the parameters \cite{Laine3}.
Perturbative analyses of the phase transition have pointed out different
mechanisms that can enhance the strength of the phase transition in the MSSM with respect to
the Standard Model case. The fact that there are additional scalar particles in the spectrum, which 
strongly couple to Standard Model fields, thus contributing significantly to  the finite-temperature effective potential opens some room for electroweak baryogenesis. The main conclusion from the perturbative analysis in the 4D theory \cite{myint}-\cite{delepine} was that
low values of the ratio of the vacuum expectation values of the two Higgs doublets $\tan\beta = {v_{2}\over v_{1}}$,  large values of the pseudoscalar mass $m_{A}$  and a very light right-handed stop were favoured.
  Given the current experimental limits on the masses in this model incorporating two-loop corrections QCD corrections is necessary in order to determine  the allowed parameter space for which  electroweak baryogenesis
can take place \cite{Espinosa}- \cite{cm}. In addition the presence of the trilinear terms in the potential is necessary in order to provide 
the required CP-violating sources for the production mechanism of electroweak
baryogenesis.
We recall that  the ratio of the vacuum expectation value of the scalar field to the temperatyre, ${\phi\over T_{c}}$, determines the rate of the sphaleron transitions in the broken phase and non-zero values for the trilinear couplings tend
to reduce the strength of the phase transition in the Higgs doublet
direction if the other parameters are kept fixed \cite{Carena1, Carena2}. This is  because the mixing in the squark sector reduces the light stop
contribution to the cubic term in the potential, it increases the critical temperature of the phase transition and  radiative corrections make the Higgs heavier,  thus weakening the strength
of the phase transition.  However,   ref. \cite{Carena2} showed that
the upper bound on the Higgs mass was arising from a zero-temperature
effect and not from the requirement of having   out-of-equilibrium sphaleron transitions. A lighter stop can compensate for the three effects that
weaken  the  phase transition.

Alternatively the analysis of the phase transition can be performed in
the context of a bosonic effective 3D theory obtained using dimensional reduction
\cite{KLRS}, \cite{ginsparg}- \cite{Nieto}. Perturbative calculations within the effective theory  reproduce in a relatively simple way the results
obtained with the 4D perturbative analysis. The additional benefit of the construction of the purely bosonic  effective
theory is that it provides a  link between the parameters employed in 3D lattice simulations with the underlying 4D parameters of the theory.
  The  parameters
in the 3D Lagrangian
are obtained using dimensional reduction at high temperature
by matching the static Green's functions in the two theories,
to a given order in the perturbative expansion, by  integrating out the non-zero Matsubara modes with
masses of the order of $\pi T$, where $T$ is the temperature. A further  reduction can also be performed noting that
 some of the static modes  in the theory have acquired thermal masses proportional to a gauge coupling multiplied by the temperature, $\sim g_{w}T, g_{s}T$.
These so-called heavy particles can then be integrated out as well.
 References \cite{KLRS},  \cite{Laine3}, \cite{Laine}, \cite{cline}-\cite{Losada3}, give more details concerning the construction of
effective theories for both  the Standard Model and the MSSM.

Interestingly the analysis of the phase transition with a light stop 
shows that for some region of parameter space a possible two-stage phase transition,
first into a colour and and charge breaking (CCB) minimum and subsequently to the
broken $SU(2)$ minimum occurs.
  Initial lattice analysis suggest
that the second transition is too strong and might not have taken place
on cosmological time scales \cite{Laine3}.
 Furthermore, in a recent paper Cline et al. \cite{Cline2} showed that although tunneling into
the CCB minimum is viable the next step of getting out of this minimum appears
not to be possible. Given this result, an accurate determination of the
critical temperatures of the phase transitions can exclude an additional region
of parameter space.  Here we closely follow the procedure of refs. \cite{KLRS, Losada3}
in order to determine very precisely
 the critical temperature of the phase transitions. In particular, the 3D scalar masses require ultraviolet renormalization and a two-loop calculation (in 4D) must be performed to fix the scales
entering the mass parameters. This requires a
full 4D effective-potential calculation incorporating mixing effects
arising from the trilinear terms in the MSSM potential. In this paper we extend the analysis of
ref. \cite{Losada3}
to identify the range of parameter space for which a sufficiently strong first-order phase transition can occur and
when a two-stage phase transition
can exist.

The perturbative component of the analysis relies  on the validity of the high temperature 
expansion.
 The value of the masses of the particles which
are integrated out will define the regime of validity of the approach. It is
interesting to consider the effects on the  parameters of the effective theory as 
the mass of the third-generation left-handed squark doublet is varied.
We present the formulae that allow us to continously pass from
the low-$m_{Q}\sim 300$ GeV to large-$m_{Q}\sim 1$ TeV regime at one-loop with squark mixing and, for the latter regime, 
we compare our with results presented in the literature.   In fact as we  will  discuss
below for values
 the most relevant effects from the trilinear couplings are one-loop effects.
For large values of $m_{Q}$, the high temperature expansion is no longer valid; in this case we estimate the two-loop effects of the trilinear couplings.

The paper is organized as follows: in section 2 we present the dimensional
reduction to the effective bosonic theory at one-loop valid for values of $m_{Q} \sim 300$ GeV.   Section 2.2 presents
a further one-loop reduction in the 3D theory,  eliminating the heavy fields. 
In section 3 we present our results for the
critical temperatures and the strength of the phase transition. The allowed region for electroweak baryogenesis to occur is also given here. Finally, in section4 we conclude. Appendix A presents the one-loop formulae needed when the trilinear terms are
included and shows how the contributions to the 3D masses and couplings are modified for
large values of third-generation left-handed squark doublet mass. In appendix B we give the expression for the two-loop unresummed effective potential in 4D, fully incorporating
mixing effects, which is necessary for evaluating $\Lambda_{H_{3}}$, $\Lambda_{U_{3}}$. The contribution from the ``heavy'' particles that were integrated out at the second stageis given in this appendix. We discuss in appendix C the relevant zero-temeperature
effects that must be included in our analysis.

\section{Dimensional Reduction}

We will now perform dimensional reduction by matching, as was previously
 done in refs. \cite{KLRS},\cite{Laine3},\cite{Laine}-\cite{Losada3}  for different models. Our initial 4D Lagrangian
corresponds to the MSSM in the large-$m_{A}$ limit.
 The particles that
contribute to the thermal bath are the Standard Model particles plus 
third-generation squarks: $\tilde{t}_{L}, \tilde{b}_{L}, \tilde{t}_{R},
 \tilde{b}_{R}$. There are two stages of reduction. The first one corresponds
to the integration out of all non-zero Matsubara modes, that is
with a thermal mass of the order of $\sim \pi T$. We calculate
all one-loop contributions  to mass terms and coupling constants of the
static fields to order $g^{4}$, where $g$ denotes a gauge or top Yukawa coupling.
The second stage of reduction corresponds to the integration of heavy particles with
masses of the order of $g_{w} T$, $g_{s} T$.

\subsection{First stage}

The potential in the 3D effective theory after integration over non-zero
Matsubara modes is of the form

\begin{eqnarray}
V &=& m_{H_{3}}^{2} H^{\dagger}H + \lambda_{H_{3}} (H^{\dagger}H)^2
+ m_{U_{3}}^{2} U^{\dagger}U + \lambda_{U_{3}} (U^{\dagger}U)^2 
+ \gamma_{3} (H^{\dagger}H)(U^{\dagger}U)\nonumber \\
&+& m_{Q_{3}}^{2} Q^{\dagger}Q  + m_{D_{3}}^{2} D^{\dagger}D 
+ \Lambda_{3}^{Q} (H^{\dagger}H)(Q^{\dagger}Q) + 
\Lambda_{4}^{c}  (H^{\dagger}Q)(Q^{\dagger}H)  \nonumber \\
&+& (\Lambda_{4}^{s} +
h_{t}^{L})|\epsilon_{ij} H^{i}Q^{j}|^{2}
+ (h_{t}^{QU} +
g_{s_{1}}^{QU})Q_{i\alpha}^{*}U_{\alpha}^{*}Q_{i\beta}U_{\beta}\nonumber \\
&+& g_{s_{2}}^{QU} U_{\alpha}U_{\alpha}^{*}Q_{j\gamma}^{*}Q_{j\gamma}
+ g_{s_{1}}^{QD} D_{\alpha}D_{\beta}^{*}Q_{j\beta}^{*}Q_{j\alpha}\nonumber \\
&+&  g_{s_{2}}^{QD} D_{\alpha}D_{\alpha}^{*}Q_{j\gamma}^{*}Q_{j\gamma}
+  g_{s_{1}}^{UD} U_{\alpha}U_{\gamma}^{*}D_{\gamma}^{*}D_{\alpha}\nonumber \\
& +& g_{s_{2}}^{UD} U_{\alpha}U_{\alpha}^{*}D_{\gamma}^{*}D_{\gamma}
+ \Lambda_{1} (Q^{\dagger}Q)^{2} +  \lambda_{D_{3}} (D^{\dagger}D)^{2}
+ \lambda_{Q_{3}}(Q_{i}^{\dagger}Q_{i})^{2}  \nonumber \\
& +& g_{s_{1}}^{QQ}
Q_{i\alpha}^{*}Q_{j\alpha}^{*}Q_{i\gamma}Q_{j\gamma}
+ g_{s_{2}}^{QQ} Q_{i\alpha}Q_{i\alpha}^{*}Q_{j\gamma}^{*}Q_{j\gamma} \nonumber \\
&+& {1\over 2} m_{A_{0}}^{2} A_{0}^{a}A_{0}^{a}  + {1\over 2} m_{C_{0}}^{2} C_{0}^{A}C_{0}^{A}
+ {1\over 4} g_{w_{3}}^{2} (H^{\dagger}H)(A_{0}^{a}A_{0}^{a}) \nonumber \\
&+& {1\over 4} g_{s_{3}}^{2} C_{0}^{A} C_{0}^{B} (U^{*})^{\dagger}\lambda^{A}\lambda^{B}U^{*}- \epsilon_{ij} \overline{X}_{t} h_{t} H^{i} Q^{j} U^{*}.
\label{3dtreepot}
\end{eqnarray}
Here $H$ is the Higgs doublet field, $U (D)$ is the right-handed stop (sbottom) field, and $Q$ is the third generation left-handed squark doublet field. The longitudinal components of the SU(2) and SU(3) gauge fields are denoted by $A_{o}$ and $C_{o}$, respectively. The Latin (Greek) indices indicate
SU(2) (SU(3)) components. As usual, the fields in eq. (\ref{3dtreepot}) are the static
components of the scalar fields properly renormalized, the dimension of the
fields in 3D is [GeV]$^{1/2}$. Quartic couplings are of order $g_{i}^{2}(h_{t}^{2}) T$, having dimensions of [GeV]; here $g_{i}(h_{t})$ denotes a gauge  (top Yukawa) coupling. The zero-temperature trilinear coupling is $X_{t} = A\sin\beta -\mu \cos\beta = \tilde{A}_{t} \sin\beta$. We work throughout in Landau gauge.
In the following we have not included the correction to the quartic coupling 
between the doublet Higgs field and the triplet scalar field $A_{0}$, or the
corresponding correction for the SU(3) counterparts.
We work throughout in the Landau gauge. We do not rewrite the
expressions for the 3D masses ($m_{A_{o}}, m_{C_{o}}$) and all of the quartic couplings which
are not modified by the presence of the trilinear couplings, they are given in ref. \cite{Losada3}.

\subsubsection{Mass terms}

For the  Higgs doublet we have\footnote{Throughout the paper we mostly neglect the hypercharge coupling $g'$. The only exception is in the contribution to the tree-level
expression of the Higgs self-coupling $\lambda$, as this latter quantity is fundamental
in determining the strength of the phase transition.} 

\begin{eqnarray}
m_{H_{3}}^{2} &=& m_{H}^{2}\biggl(1 + {9\over 4} g_{w}^{2} {L_{b}\over 16\pi^{2}}
 - 3 h_{t}^{2}\sin^{2}\beta {L_{f}
\over 16\pi^{2}}\biggr)\nonumber \\
 &+&T^{2}\biggl({\lambda\over 2} + {3\over 16} g_{w}^{2}
 + {1\over 16} g'^{2} 
+ {1\over 4} h_{t}^{2}\sin^{2}\beta + {1\over 4}
(2h_{t}^{2}\sin^{2}\beta + 2\lambda_{3} + \lambda_{4})\biggr)\nonumber\\
&-&{L_{b}\over 16\pi^{2}}\biggl(6 \lambda m_{H}^{2}+ 3(m_{Q}^{2} + m_{U}^{2})h_{t}^{2}\sin^{2}\beta + N_{c} h_{t}^{2} X_{t}^{2}\biggr),
\label{mmH3}
\end{eqnarray}
where the Higgs mass parameter is denoted by $m_{H}$,
 and $\lambda = {(g_{w}^{2} + g'^{2})\over 8}\cos^2 2\beta, \lambda_{3} = {g_{w}^{2} \over 4}, \lambda_{4} = -{g_{w}^{2}\over 2}$.  Similarly, for the third-generation squark mass terms we have

\begin{eqnarray}
m_{U_{3}}^{2} &=& m_{U}^{2}\biggl(1 + 4 g_{s}^{2} {L_{b}\over 16\pi^{2}}\biggr) +
 T^{2} \biggl({1\over 3} g_{s}^{2}
 + {2\over 3} \lambda_{U} + {1\over 6} h_{t}^{2}\sin^{2}\beta +
{1\over 6}h_{t}^{2} \biggr)
\nonumber \\
&-& {L_{b}\over 16 \pi^{2}}\biggl({4\over 3} g_{s}^{2} m_{U}^{2} + 2 h_{t}^{2} \sin^{2}\beta( m_{H}^{2}+
m_{Q}^{2})  + 2 h_{t}^{2} X_{t}^{2}\biggr),
\label{mmU3}
\end{eqnarray}

\begin{eqnarray}
m_{Q_{3}}^{2}& =& m_{Q}^{2}\biggl(1 + ({9\over 4} g_{w}^{2} 
 + 4 g_{s}^{2}){L_{b}\over 16\pi^{2}}\biggr) + T^{2}\biggl({3\over 16} g_{w}^{2}
+{ \lambda_{1}\over 2} + {4\over 9} g_{s}^{2}
 + {1\over 12} h_{t}^{2}(1 + \sin^{2}\beta)\biggr) \nonumber \\
&-& {L_{b}\over 16\pi^{2}}\biggl({4\over 3} g_{s}^{2} m_{Q}^{2} + 6 \lambda_{1} m_{Q}^{2}
+ h_{t}^{2} m_{U}^{2} + h_{t}^{2}\sin^{2}\beta m_{H}^{2} + h_{t}^{2} X_{t}^{2}\biggr),
\label{mQ3}
\end{eqnarray}
where the soft SUSY-breaking mass for the third generation left-handed  squark
doublet is denoted by $m_{Q}$, and $\lambda_{U} = {g_{s}^{2}\over 6}$, $\lambda_{1} = {g_{w}^{2}\over 8}$.



\subsubsection{Couplings}

 We present here the expressions for the trilinear couplings in 3D, which were
absent for the case of zero-squark mixing. The modifications
of the scalar quartic couplings are suppressed at this stage for the
values of the mixing parameters which we will consider, and we do not
include the corrections from diagrams of those in fig. 1.  However,
at the next-stage of integration-out we will include them:




\begin{eqnarray}
\overline{X}_{t} h_{t} & = & X_{t}h_{t}T^{1/2}\biggl(1 + {9\over 4} g^2 {L_{b}\over 16\pi^2} - 
 {N_{c}\over 2}h_{t}^{2}\sin^{2}\beta  
{L_{f}\over 16\pi^2}\nonumber \\
&+& 4g_{s}^{2} {L_{b}\over 16\pi^2} \biggr)
- X_{t}h_{t}T^{1/2}\biggl(\lambda_{3} + 2\lambda_{4}\sin^{2}\beta + 3h_{t}^{2}\sin^{2}\beta
+ N_{c} h_{t}^{2}  \nonumber \\
& -& {4\over 3} g_{s}^{2} \biggr){L_{b}\over 16\pi^2}.
\label{Xfu}
\end{eqnarray}

\subsection{Second stage}

 Another
simplification of the effective theory can be obtained by integrating out the
scalar fields, which are massive at the transition point.
As we have seen, the static modes corresponding to the scalar fields $Q, D, A_{o}, C_{o}$ acquired thermal masses proportional to $\sim g_{w(s)}T$, as a consequence of the integration out of the non-zero Matsubara modes.
The second stage proceeds in exactly the same way as in refs. \cite{Laine2, Losada3}.
We include the additional corrections arising from the couplings we have considered.

\subsubsection{Couplings}
The final expression for the tree-level 3D potential is given by

\begin{equation}
V_{3D} = \overline{m}_{H_{3}}^{2} H^{\dagger}H + \overline{\lambda}_{H_{3}}(H^{\dagger}H)^{2} + \overline{m}_{U_{3}}^{2} U^{\dagger}U + \overline{\lambda}_{U_{3}}(U^{\dagger}U)^{2}
+ \overline{\gamma}_{3} H^{\dagger}H U^{\dagger}U,
\label{v3D}
\end{equation}
where the scalar couplings are now

\begin{eqnarray}
\overline{\lambda}_{H_{3}} &=& \lambda_{H_{3}} - {3\over 16} {g_{w_{3}}^{4}\over 8\pi m_{A_{0}}}
-{3\over 8\pi m_{Q_{3}}}\biggl(\Lambda_{3}^{2} + \Lambda_{3}(\Lambda_{4}^{c} + 
\Lambda_{4}^{s}) \nonumber \\
&+& {1\over 2} \biggl((\Lambda_{4}^{c})^{2} + (\Lambda_{4}^{s})^{2}\biggr) + h_{t}^{L} \Lambda_{3}^{Q}
+ h_{t}^{L}\Lambda_{4}^{s} + {1\over 2}( h_{t}^{L})^{2}\biggr)\nonumber \\
&+& 3\overline{X}_{t}^{2} h_{t}^{2}(\Lambda_{3}^{Q} + \Lambda_{4}^{s} +
h_{t}^{L}){1\over8\pi m_{Q_{3}}}{1\over (m_{Q_{3}} + m_{U_{3}})^{2}} \nonumber \\
&+& 3\overline{X}_{t}^{2} h_{t}^{2} \gamma_{3}{1\over8\pi m_{U_{3}}}{1\over (m_{Q_{3}} + m_{U_{3}})^{2}} - {3\over 2}\overline{X}_{t}^{4} h_{t}^{4} f(m_{Q_{3}},m_{U_{3}},m_{U_{3}}),
\label{lamH3}
\end{eqnarray}

\begin{eqnarray}
\overline{\lambda}_{U_{3}} &=& \lambda_{U_{3}} -
 {13\over 36} {g_{s_{3}}^{4}\over 8\pi m_{C_{0}}}
-{1\over 8\pi m_{D_{3}}}\biggl({1\over 2}(g_{s_{1}}^{UD} + g_{s_{2}}^{UD})^{2}
 + (g_{s_{2}}^{UD})^{2}\biggr) \nonumber \\
&-& {1\over 8\pi m_{Q_{3}}}\biggl((h_{t}^{QU})^{2} - 2 h_{t}^{QU} g_{s_{1}}^{QU} + 
2h_{t}^{QU} g_{s_{2}}^{QU} +(g_{s_{1}}^{UD} + g_{s_{2}}^{UD})^{2} +
2(g_{s_{2}}^{UD})^{2}\biggr) \nonumber \\
&+& \overline{X}_{t}^{2} h_{t}^{2}(h_{t}^{QU} + g_{s_{1}}^{QU} + g_{s_{2}}^{QU}){1\over8\pi m_{Q_{3}}}{1\over (m_{Q_{3}} + m_{H_{3}})^{2}} + 
\overline{X}_{t}^{2} h_{t}^{2} \gamma_{3} {1\over8\pi m_{H_{3}}}{1\over (m_{Q_{3}} + m_{H_{3}})^{2}}\nonumber \\
&-& \overline{X}_{t}^{4} h_{t}^{4} f(m_{Q_{3}},m_{H_{3}},m_{H_{3}}),
\label{lamU3}
\end{eqnarray}

\begin{eqnarray}
\overline{\gamma_{3}} &=& \gamma_{3} -{1\over 8\pi m_{Q_{3}}}(h_{t}^{QU} + g_{s_{1}}^{QU}+ 3 g_{s_{2}}^{QU})(2\Lambda_{3}^{Q}
+ \Lambda_{4}^{c} + \Lambda_{4}^{s} + h_{t}^{L}) \nonumber \\
&+& \overline{X}_{t}^{2} h_{t}^{2}(\Lambda_{3}^{Q} + {1\over 2}(\Lambda_{4}^{s} + h_{t}^{L})){1\over8\pi m_{Q_{3}}}{1\over (m_{Q_{3}} + m_{H_{3}})^{2}}\nonumber \\
&+& \overline{X}_{t}^{2} h_{t}^{2} {1\over 2}( h_{t}^{QU} + g_{s_{1}}^{QU} + 3 g_{s_{2}}^{QU}){1\over8\pi m_{Q_{3}}}{1\over (m_{Q_{3}} + m_{U_{3}})^{2}} \nonumber \\
&+&
3 \overline{X}_{t}^{2} h_{t}^{2} \Lambda_{H_{3}}{1\over8\pi m_{H_{3}}}{1\over (m_{Q_{3}} + m_{H_{3}})^{2}} + 4\overline{X}_{t}^{2} h_{t}^{2} \Lambda_{U_{3}}{1\over8\pi m_{U_{3}}}{1\over (m_{Q_{3}} + m_{U_{3}})^{2}} \nonumber \\
&+& \overline{X}_{t}^{2} h_{t}^{2} \gamma_{3} {1\over 4\pi}{1\over (m_{Q_{3}} + m_{U_{3}})}{1\over (m_{Q_{3}} + m_{H_{3}})}{1\over (m_{U_{3}} + m_{H_{3}})} - \overline{X}_{t}^{4} h_{t}^{4} f(m_{Q_{3}},m_{H_{3}},m_{U_{3}}) .
\label{gamma3}
\end{eqnarray}

and

\begin{equation}
f(m_{1},m_{2},m_{3}) = {1\over 8\pi} {2m_{1} + m_{2} + m_{3} \over
m_{1}(m_{1} + m_{3})^{2} (m_{1} + m_{2})^{2} (m_{2} + m_{3})}.
\end{equation}

\subsubsection{Mass terms}

The one-loop contribution to the mass terms can be obtained directly as shown in
ref. \cite{KLRS}:

\begin{eqnarray}
\overline{m}_{H_{3}}^{2} &=& m_{H_{3}}^{2}\biggl(1 - {N_{c}h_{t}^{2}\overline{X}_{t}^{2}\over 12\pi(m_{Q_{3}} +
m_{U_{3}})^{3}}\biggr) - {3\over 16\pi} g_{w_{3}} m_{A_{0}} -{3\over 4\pi}
(2\Lambda_{3}^{Q} +  \Lambda_{4}^{c} + \Lambda_{4}^{s} + h_{t}^{L})m_{Q_{3}} \nonumber \\
&-& 3 h_{t}^{2}\overline{X}_{t}^{2}{1\over 4\pi(m_{Q_{3}} +
m_{U_{3}})},
\label{mH3}
\end{eqnarray}

\begin{eqnarray}
\overline{m}_{U_{3}}^{2} &=& m_{U_{3}}^{2}\biggl(1 - {2 h_{t}^{2}\overline{X}_{t}^{2}\over 12\pi(m_{Q_{3}} +
m_{H_{3}})^{3}}\biggr) - {1\over 3\pi} g_{s_{3}} m_{C_{0}} -{1\over 4\pi}
(2 h_{t}^{QU} + 2 g_{s_{1}}^{QU} + 6 g_{s_{2}}^{QU})m_{Q_{3}}\nonumber \\
& -&{1\over 4\pi}(2g_{s_{1}}^{UD}
+ 6g_{s_{2}}^{UD})m_{D_{3}} -  2 h_{t}^{2}\overline{X}_{t}^{2}{1\over 4\pi(m_{Q_{3}} +
m_{H_{3}})}.
\label{mU3}
\end{eqnarray}

In order to precisely fix the scales of the
couplings that appear in the thermal polarizations of eqs. (2) and (3), one needs to perform 
a 2-loop evaluation of the effective potential.
 In addition, the mass parameters are
renormalized in the 3D theory:

\begin{equation}
\overline{m}_{H_{3}}^{2}(\mu) = \overline{m}_{H_{3}}^{2} + {1\over (16\pi^{2})} f_{2m_{H}}\log{\Lambda_{H_{3}}\over \mu},
\label{mH3mu2}
\end{equation}

\begin{equation}
\overline{m}_{U_{3}}^{2}(\mu) = \overline{m}_{U_{3}}^{2} + {1\over (16\pi^{2})} f_{2m_{U}}\log{\Lambda_{U_{3}}\over \mu}.
\label{mU3mu2}
\end{equation}
The expressions for the two-loop beta functions $f_{2m_{H}},f_{2m_{U}}$ for the mass parameters
have been given in ref. \cite{Laine2}. 
 In appendix B we perform a two-loop calculation
of the effective potential for the  $H$ ($\phi$-direction) and $U$($\chi$-direction) fields, including the
effects of mixing with the third generation left-handed squark doublet\footnote{ Some of the
results given previously in the literature are corrected.}. We incorporate all of the
corrections to the 3D couplings obtained in the previous sections, so as to determine the exact values of
$\Lambda_{H_{3}}$ and $\Lambda_{U_{3}}$. Using the results of
appendix B it can be checked that the trilinear couplings do not produce     scale dependent logarithmic
contributions  to the mass terms. In other words when both stop fields are light, the residual dependence on the mixing
term in  the divergent parts of the contributions to  the mass terms appears only through the couplings which enter the expressions for the beta functions $f_{2m_{H}},f_{2m_{U}}$. These couplings are given by the one-loop expressions in  equations (\ref{lamH3}) -(\ref{gamma3}) .
We will analyse the effect of including these corrections on the critical temperatures \footnote{ The necessary modifications of $f_{2m_{H}},f_{2m_{U}}$ in the case of non-zero mixing for large values of $m_{Q}$ appear in ref. \cite{Carena2}. The results
of appendix B are no longer applicable as the high temperature expansion
expressions for the $D$ functions are not valid for the third generation
left-handed squark doublet, see below.}.

\section{Results}

With  the results of the previous sections and those in  appendices B and C we can analyse the phase transition for values of $m_{Q} = 300$ GeV.
 In fig. \ref{Tcmt1} we
show the critical temperatures for the transitions in the $\phi$ (solid)- and
$\chi$ (dotted)-directions as a function of the lightest stop  mass $m_{\tilde{t}_{2}}$, for $\tan\beta =5$ and for
values of the parameter $\tilde{A}_{t} = {X_{t}\over \sin\beta} = 100, 200$ GeV. We find that, for  $m_{Q} \sim 300$ GeV, there
 still is a region in which a two-stage phase transition can occur. This region
is to the left of the crossing point of the curves. As shown in figs. \ref{Tcmt1} and \ref{vTmt1},  the range of values of the lighest stop mass for which
the phase transition is sufficiently strong and a two-stage phase transition occurs decreases as $X_{t}$ increases. Thus the experimental
bound on the stop mass puts an
upper bound on the value  of the mixing parameter. The figures display  the effects produced by the  non-zero trilinear couplings as  described in the introduction. Keeping all other parameters fixed, lowering the
stop mass will decrease the critical temperature in the $\phi$-direction, slightly reduce
the Higgs mass and strengthens the transition. The light-stop mechanism of enhancing the
phase transition also eliminates the restriction of having low values of $\tan\beta$.

The allowed region in parameter space is shown in fig. \ref{comp1}, as a function of  the Higgs and stop masses, for  $\tilde{A}_{t} = 200$ GeV, the region on the left of the solid line indicates
when a sufficiently strong first-order phase transition occurs. The
dotted line gives the condition for absolute stability of the
physical vacuum.  As explained in appendix C, to the left of this line the colour-breaking minimum is
lower than the physical one at zero temperature. The dashed line is obtained
when the critical temperatures of the transitions in the $\phi$- and $\chi$-directions
are the same. A two-stage phase transition occurs to the left of the dashed line. If the second transition from the CCB minimum  to the electroweak minimum
does not occur, as indicated by the analysis of ref. \cite{Cline2}, there are stronger constraints on the allowed region. For each
value of the mixing parameter $\tilde{A}_{t}$ the available region 
is restricted to the band within the dashed and the solid lines.
Note that,  for low values  of $m_{Q}$  and
non-zero
squark mixing, there is no cross-over
 between the dashed and dotted lines. The end-points of the lines correspond to the maximum value of the Higgs mass which is reached by the  effect of the zero-temperature radiative
corrections for a given value of $m_{Q}$ and $\tilde{A}_{t}$.
In fig. \ref{vteffec} we display the full allowed region in parameter space coming from
the requirement of having a strong enough phase transition. The solid vertical line is determined t varying $\tan\beta$ for $\tilde{A}_{t}=0$ and the solid diagonal line corresponds to
the variation of $\tilde{A}_{t}$ for $\tan\beta=12$. The region to the left and below
these solid lines provides a sufficiently strong phase transition for 
values of $2 \lsi \tan\beta \lsi 12$ and $0 \lsi \tilde{A}_{t} \lsi 280$ GeV. The dashed line shown defines when the critical temperatures in the $\phi$- and
$\chi$-directions are equal for the same variations of $\tan\beta$ and
 $\tilde{A}_{t}$. For this value of $m_{Q}$ the maximum allowed masses
are $m_{h} \lsi 105$GeV and $m_{\tilde{t}_{2}} \lsi 170$GeV.

We now turn to the analysis of the phase transition for $m_{Q}=1$ TeV. The expressions
for the 3D parameters are given in  appendix A. The one-loop
calculation is exact, the two-loop contributions are only estimated, using the
corresponding two-loop beta functions for the mass parameters with non-zero mixing at
large $m_{Q}$ given in ref. \cite{Carena3}. Here the scales $\Lambda_{H_{3}}$ and $\Lambda_{U_{3}}$ have been
fixed to agree with the numerical results obtained in the case with zero-mixing in ref. \cite{Losada3}.
We point out that  the  exact contributions from the two-loop  sunset and figure eight diagrams when the high temperature expansion is no longer valid  are not known \footnote{We hope to
return to this point in the future.}. These contributions  are necessary for terms involving  the third generation
left-handed squark doublet for large values of $m_{Q}$. Figure \ref{TcmQlarge} shows that the phase-diagram structure is maintained 
for $m_{Q} = 1$ TeV for several values of $\tilde{A}_{t}$. Again, a possible two-stage phase
transition persists, in agreement with previous results \cite{Carena2, Seco}. We remark
that the qualitative effects of increasing  $\tilde{A}_{t}$ remain unchanged with respect to the case of $m_{Q}=300$GeV. In figure \ref{comp2}  the solid line determines the
constant ratio ${\phi\over T_{c}} =1$,  as a function
of the lightest stop mass for $\tan\beta=5$, varying the mixing parameter
in the range $0 \lsi \tilde{A}_{t} \lsi 650$ GeV. The dashed line is obtained when the critical temperatures in the $\phi$- and $\chi$- directions
are the same.  The dotted line  gives the zero-temperature condition
for absolute stability of the physical minimum. Imposing the constraint  which eliminatesthe region where tunneling into a  
CCB minima at finite temperature is possible reduces the allowed region to  the band  between the dashed and solid line. The experimental  constraints on the stop mass $m_{\tilde{t}_{2}} \gsi 80$ GeV, will also restrict the value of the mixing parameter. Finally, in  fig. \ref{param2} we present
the contours of  ${\phi\over T_{c}} =1$ in the $m_{h}$-$m_{\tilde{t}_{2}}$ plane varying
$\tan\beta$ for $\tilde{A}_{t}= 0,100,200,300,400,600$. The regions of possible two-stage phase transitions  defined by each set of values of $A_{t}$ and $\tan\beta$ are
not shown. The maximum allowed Higgs mass  is now $m_{h} \lsi 120 $ GeV.

\section{Conclusions}
We have performed a full two-loop dimensional reduction
of 4D MSSM parameters to the 3D couplings and masses of the effective theory, including
the effects from the trilinear terms in the potential.
 The preservation
of the baryon asymmetry can be ensured tuning the stop mass
in order to compensate the negative effects of the mixing term
on the strength of the phase transition.  We conclude that the
 direct stop searches 
restrict the amount of mixing in the stop sector. The allowed range of
masses is $m_{h} \lsi 107$ GeV and $m_{\tilde{t}_{1}} \lsi 170$ GeV for $m_{Q}=300$ GeV. Our calculation allows us to determine the critical temperatures of
the transitions precisely. We find that the phase-diagram still allows a possible two-stage phase transition for a small
range of values of $m_{\tilde{t}_{2}}$ and any value of $\tilde{A}_{t}$ for values
of $m_{Q}$ for which the high-temperature expansion can be applied. This range of values is shifted
for different values of the stop mixing parameter.
If the second phase transition does not occur there is
a further restriction of the allowed regions of parameter space for electroweak baryogenesis for each value of the mixing parameter and $\tan\beta$.
 At large values of $m_{Q}$ the qualitative dependence on the parameters
in the theory remains unchanged.

\appendix

\section{One-loop contributions from trilinear terms and large-$m_{Q}$}

The first diagram that contributes to the two-point functions  corresponds to fig. 1a  \cite{dolan,funakubo}:

\begin{equation}
I(m) = A(m) + f_{1}(a)
\label{A1}
\end{equation}
where

\begin{eqnarray}
A(m) &=& -{m^{2} \over 16\pi^{2}}\biggl(\Delta + \ln{\mu^2\over m^2} + 1\biggr)\nonumber \\
f_{1}(a) &=& {T^2\over 2\pi^2 }\int_o^{\infty} dx {x^2\over (x^2 + a^2)^{1\over 2}} {1\over e^{(x^2 + a^2)^{1\over 2}} -1}
\label{Af1}
\end{eqnarray}
and $a= {m\over T}$.
The  expression that is valid for a  high-temperature expansion is

\begin{equation}
I(m) = {T^2\over 12}(1 + \epsilon i_{\epsilon}) - {m^2\over 16\pi^2}\biggl({1\over \epsilon} +L_{b}\biggr) + O({m^{2}\over T^{2}}),
\label{IM}
\end{equation}
where $L_{b}=2 \log{\overline{\mu} e^{\gamma}\over 4\pi T} \approx 2 \log{\mu\over 7.055 T}$, $L_{f} = L_{b} + 4\log2$. Here $\overline{\mu}$ is the mass scale
defined by the modified minimal subtraction ($\overline{MS}$) scheme. For a low-temperature expansion that is $m \gg T$ only the zero-temperature contribution remains as the temperature-dependent part is suppressed. Trilinear coupling induce an additional contribution to the two-point function shown in fig. 1b:

\begin{equation}
I_{2}(m_{1},m_{2}) = B_{o}(p,m_{1},m_{2}) + {1\over 2\pi^2} f_{2}(a,b),
\label{Bf2}
\end{equation}
where $ B_{o}(p,m_{1},m_{2})$ is the usual Veltman-Passarino scalar function and
in particular,

\begin{eqnarray}
 B_{o}(0,m_{1},m_{2})& =& {1\over 16\pi^2}\biggl(\Delta + 1 + \ln{\mu^2\over m_{2}^{2}} + \biggl({m_{1}^{2}\over m_{1}^{2} - m_{2}^{2}}\biggr) \ln{m_{2}^{2}\over m_{1}^{2}}\biggr) \nonumber \\
f_{2}(a,b) &=& -{1\over a^2-b^2}\int_o^{\infty} dx\biggl( {x^2\over (x^2 + a^2)^{1\over 2}} {1\over e^{(x^2 + a^2)^{1\over 2}} -1} -{x^2\over (x^2 + b^2)^{1\over 2}} {1\over e^{(x^2 + b^2)^{1\over 2}} -1}\biggr)
\label{f2ab}
\end{eqnarray}
where $a=m_{1}/T, b= m_{2}/T$.

For a high-temperature expansion this becomes

\begin{equation}
I_{2}(m_{1},m_{2}) =  {1\over 16\pi^2}({1\over \epsilon} +L_{b}) + O({m^{2}\over T^{2}}) + O({p^{2}\over T^{2}}).
\label{A2}
\end{equation}

For $m_{1}\ll T$ and $m_{2}\gg 2\pi T$ is

\begin{equation}
I_{2}(m_{1},m_{2}) =   B_{o}(p,m_{1},m_{2}) +{1\over m_{1}^2} f_{1}(m_{2})
\label{lTf2}
\end{equation} 
and for the case in which $m_{1}=m_{2}\gg T$ this reduces to 
the zero-temperature contribution

\begin{equation}
I_{2}(m,m) =  B_{o}(0,m,m) = {1\over 16\pi^2} \ln{\mu^2\over m^2}
\label{Bmm}
\end{equation}
plus wave-function renormalization effects.
We remind the reader that this can be obtained from
$I_{2}(m,m) = -{\partial\over \partial m^2}$ I(m). Similarly, we can obtain
the corresponding results for other diagrams, in fig. 1 taking the appropiate
derivatives. We now give only the low-temperature expressions
valid for $m_{1}\gg T$, the corresponding high-temperature expressions can be found
in the literature:

\begin{equation}
I_{3}(m_{1},m_{2},m_{2}) = C_{o}(0,m_{1},m_{2},m_{2}) + {1\over 16\pi^2}{1\over m_{1}^2}
\ln{m_{2}^2\over \mu_{T}^{2}} 
 \label{Cmm1}
\end{equation}

\begin{equation}
I_{3}(m_{1},m_{1},m_{2}) = C_{o}(0,m_{1},m_{1},m_{2}) + {1\over m_{1}^4}f_{1}(m_{2})
 \label{Cmm2}
\end{equation}

\begin{equation}
I_{4}(m_{1},m_{1},m_{2},m_{2}) = D_{o}(0,m_{1},m_{1},m_{2},m_{2}) + {1\over m_{1}^4}{1\over 16\pi^2} \ln{m_{2}^{2}\over \mu_{T}^2}.
 \label{Cmm3}
\end{equation}
where the functions  $C_{o}$ and $D_{o}$ are the scalar one-loop functions
of Veltman-Passarino. 

The  3D couplings   in equation \ref{v3D} for the case of  large $m_{Q}$ are
 given by:

\begin{eqnarray}
\overline{\lambda}_{H_{3}} &=& \lambda T\biggl(1 + {9\over 2} g_{w}^{2}{ L_{b}\over 16\pi^{2}} -
6 h_{t}^{2} \sin^{2}\beta {L_{f}\over 16 \pi^2}\biggr) \nonumber \\
&-&{T\over 16 \pi^{2}} \biggl[ L_{b}\biggl( {9\over 16}g_{w}^4 + 
12 \lambda^{2}  + {1\over 2} h_{t}^{4}\sin^{4}\beta \biggr)+ 3\biggl( \lambda_{3}^{2} + \lambda_{3}\lambda_{4}
 + \lambda_{4}^{2}\cos^{4}\beta
+ \lambda_{4}^{2}\sin^{4}\beta  \nonumber \\
&+& h_{t}^{2}\sin^{2}\beta\bigl( \lambda_{3} + \lambda_{4} \sin^{2}\beta \bigr)
 + {1\over 2}h_{t}^{4}\sin^{4}\beta \biggr) \ln{\mu^2\over m_{Q}^{2}}  
+ {3\over 8}g_{w}^{4} 
+ 3 h_{t}^{4}\sin^{4}\beta L_{f} \nonumber \\
&+& 3 X_{t}^{2} h_{t}^{2} (\lambda_{3} + \lambda_{4}\sin^2\beta + h_{t}^{2} \sin^{2}\beta){1\over m_{Q}^{2}}
\nonumber \\
&+&  3 X_{t}^2 h_{t}^{4} \sin^{2}\beta (-{1\over m_{Q}^{2}}(1- \ln{m_{Q}^{2}\over \mu_{T}^{2}})) - {3\over 2} X_{t}^{4} h_{t}^{4} (-{1\over m_{Q}^{4}}(2 + \ln{\mu_{T}^{2}\over m_{Q}^{2}})) - N_{c} \lambda h_{t}^{2} {X_{t}^{2}\over m_{Q}^{2}} \biggr],
\label{lalamH3}
\end{eqnarray}

\begin{eqnarray}
\overline{\lambda}_{U_{3}} &=& {g_{s}^{2}\over 6} T\biggl(1 + 8 g_{s}^{2} {L_{b}\over
16\pi^{2}}\biggr)
-{T\over
16\pi^{2}}\biggl[L_{b}\biggl({17\over 36} g_{s}^{4} + {13\over 12} g_{s}^{4} + h_{t}^{4}\biggr)  - \biggl({2\over 3}
 h_{t}^{2}g_{s}^{4}\nonumber \\
& +& {g_{s}^{2}\over 6} +
h_{t}^{4}\sin^{4}\beta\biggr) \ln{\mu^{2}\over m_{Q}^{2}} + {13\over 18} g_{s}^{4} + 2 X_{t}^{2} h_{t}^{2} (h_{t}^{2} - {1\over 3} g_{s}^{2}){1\over m_{Q}^{2}}) \nonumber \\
&+& 2 X_{t}^{2} h_{t}^{4} \sin^{2}\beta(-{1\over m_{Q}^{2}}(1- \ln{m_{Q}^{2}\over \mu_{T}^{2}}) - X_{t}^{4} h_{t}^{4} (-{1\over m_{Q}^{4}}(2 - \ln{m_{Q}^{2}\over \mu_{T}^{2}})) - {g_{s}^{2}\over 3} h_{t}^{2} {X_{t}^{2}\over m_{Q}^{2}}  \biggr],
\label{lalamU3}
\end{eqnarray}

\begin{eqnarray}
\overline{\gamma}_{3} &= &h_{t}^{2}\sin^{2}\beta T\biggl(1+ {9\over 4} g_{w}^{2} {L_{b}\over 16\pi^{2}}
 -3 h_{t}^{2}\sin^{2} 
{L_{f}\over 16\pi^{2}} + 4 g_{s}^{2} {L_{b}\over 16\pi^{2}}\biggr) \nonumber \\
&-&{T\over 16\pi^{2}}\biggl[L_{b}\biggl({4\over 3} h_{t}^{2} \sin^{2}\beta g_{s}^{2}
+ 2 h_{t}^{4}\sin^{4}\beta + 6 \lambda h_{t}^{2}\sin^{2}\beta\biggr)\nonumber \\
&+& \biggl( h_{t}^{2}(2\lambda_{3} +
\lambda_{4} + h_{t}^{2}\sin^{2}\beta\biggr) \ln{\mu^{2}\over m_{Q}^{2}}
+ X_{t}^{2} h_{t}^{2}\biggl( 2(\lambda_{3} + {1\over 2} \lambda_{4}\sin^{2}\beta + h_{t}^{2}\sin^{2}\beta + h_{t}^{2}){1\over m_{Q}^{2}} + (6 \lambda + 4 {g_{s}^{2}\over 3} \nonumber \\
&+& 2 h_{t}^{2}\sin^{2}\beta)(-{1\over m_{Q}^{2}}(1- \ln{m_{Q}^{2}\over \mu_{T}^{2}})) - X_{t}^{2} h_{t}^{2}(-{1\over m_{Q}^{4}}(2 + \ln{\mu_{T}\over m_{Q}^{2}})) - h_{t}^{2} \sin^{2}\beta ({1\over m_{Q}^{2}} + {3\over m_{Q}^{2}})  \biggr],
\label{gagamma3}
\end{eqnarray}

and the 3D masses are given by
\begin{eqnarray}
m_{H_{3}}^{2} &=& m_{H}^{2}\biggl(1 + {9\over 4} g_{w}^{2} {L_{b}\over 16\pi^{2}}
 - 3 h_{t}^{2}\sin^{2}\beta {L_{f}
\over 16\pi^{2}}\biggr)\nonumber \\
 &+&T^{2}\biggl({\lambda\over 2} + {3\over 16} g_{w}^{2}
 + {1\over 16} g'^{2} 
+ {1\over 4} h_{t}^{2}\sin^{2}\beta + {1\over 4}h_{t}^{2}\sin^{2}\beta\biggr)
- 3(h_{t}^{2}\sin^{2}\beta + 2\lambda_{3} + \lambda_{4}){m_{Q}^{2}\over 16 \pi^{2}}\biggl(\ln{\mu^{2}\over m_{Q}^{2}} + 1 \biggr)\nonumber\\
&-&{L_{b}\over 16\pi^{2}}\biggl(6 \lambda m_{H}^{2}+ 3 m_{U}^{2})h_{t}^{2}\sin^{2}\beta \biggr) - 3 {h_{t}^{2} X_{t}^{2} \over 16 \pi^{2}}\biggl( -{m_{H}^{2} \over 2 m_{Q}^{2}} + {m_{U}^{2}\over m_{Q}^{2}} \biggl(\ln{\mu_{T}^{2}\over m_{Q}^{2}} + 1\biggr) \nonumber \\
&+& \ln{\mu^{2}\over m_{Q}^{2}} +1\biggr) - 3h_{t}^{2} {X_{t}^{2}\over m_{Q}^{2}} {T^{2}\over 12}  - {3\over 16\pi} g_{w_{3}} m_{A_{0}}+ {1\over (16\pi^{2})} f_{2m_{H}}\log{\Lambda_{H_{3}}\over \mu},
\label{mmH3}
\end{eqnarray}

\begin{eqnarray}
m_{U_{3}}^{2} &=& m_{U}^{2}\biggl(1 + 4 g_{s}^{2} {L_{b}\over 16\pi^{2}}\biggr) +
 T^{2} \biggl({1\over 3} g_{s}^{2}
 + {2\over 3} \lambda_{U} + {1\over 6} h_{t}^{2}\sin^{2}\beta\biggr) -
2 h_{t}^{2} {m_{Q}^{2}\over 16 \pi^{2}}\biggl(\ln{\mu^{2}\over m_{Q}^{2}} + 1 \biggr)
\nonumber \\
&-& {L_{b}\over 16 \pi^{2}}\biggl({4\over 3} g_{s}^{2} m_{U}^{2} + 2 h_{t}^{2} \sin^{2}\beta( m_{H}^{2}+
m_{Q}^{2})\biggr)  - 2 {h_{t}^{2} X_{t}^{2}\over 16\pi^{2}}\biggl( -{m_{U}^{2}\over 2m_{Q}^{2}} \nonumber \\
& +& \ln{\mu^{2}\over m_{Q}^{2}} +1 + {m_{H}^{2}\over m_{Q}^{2}}\biggl( \ln{ \mu_{T}^{2}\over m_{Q}^{2}} + 1\biggr)\biggl) - 2 h_{t}^{2} {X_{t}^{2}\over m_{Q}^{2}} {T^{2}\over 12}  - {1\over 3\pi} g_{s_{3}} m_{C_{0}} + {1\over (16\pi^{2})} f_{2m_{U}}\ln{\Lambda_{U_{3}}\over \mu},
\label{mmU3}
\end{eqnarray}
where $\mu_{T} = 4 \pi e^{-\gamma} T$.

\section{Two-loop contributions with non-zero squark mixing}

The strategy we employ follows that of refs. \cite{KLRS,Losada3}.
The idea is to use the 4D two-loop effective potential in order to fix
the scales in the 3D theory, and to use the 3D effective potential expressions
for the Higgs and stop fields
given in ref. \cite{Laine2} to analyse the phase transition.
We  calculate the unresummed two-loop effective potential
in order to include all 4D corrections to the mass parameters; resummation
is automatically  included in the calculation of the two-loop effective
potential in the 3D theory. We must also include the
contributions  to the two-loop effective potential of the static modes, which have been
integrated out  at the second stage  (includes the effects of resummation of the heavy fields).

There are several effects that must be considered in order to obtain
all of the contributions (constant and logarithmic) to the mass 
parameters. From the 4D effective potential, one finds the two-loop contributions
from the gauge bosons, Higgs, right-handed stop, left-handed squark doublet, right-handed sbottom,
and top quark. The main difference is that the $D$ functions appearing below
correspond to the unresummed
expressions. Additionally we
must include the effects arising  at the second
stage of reduction from  the left-handed squark doublet, the right-handed sbottom, the scalar  triplet and the scalar 
octet\footnote{The expression given for the 4D effective potential
would correspond to the
usual resummed two-loop 4D effective potential if we used 
the resummed expressions in the $D$ functions appearing below.}.

We now derive  the effective potential at finite temperature using  the
background fields  $\phi$ and $\chi= \tilde{t}_{R\alpha}u^{\alpha}$, where
we have chosen the unit vector in colour space $u^{\alpha}=(1,0,0)$. 
We first write the expressions in the shifted theory of
 the mass spectrum after the first stage of integration. 

\subsection{$\phi$-direction}
The gauge-boson masses are
\begin{equation}
m_{W,Z}^{2} = {1\over 4} g_{w}^{2} \phi^{2}.
\label{mW}
\end{equation}

The stop mass matrix elements are given by

\begin{equation}
m_{\tilde{t}_{L}}^{2} (\phi) = m_{Q}^{2} + m_{t}^{2}(\phi)
+ {1\over 4}({1\over 2} - {2\over 3} \sin^{2}\theta_{W}) (g_{w}^{2} + g'^{2})\phi^{2}\cos 2\beta
\label{mtL2}
\end{equation}

\begin{equation}
m_{\tilde{t}_{R}}^{2} (\phi) = m_{U}^{2} + m_{t}^{2}(\phi)
+ {1\over 4}({2\over 3} \sin^{2}\theta_{W}) (g_{w}^{2} + g'^{2})\phi^{2}\cos 2\beta
\label{mtL2}
\end{equation}

\begin{equation}
m_{\tilde{t}_{LR}}^{2} (\phi)= {h_{t}\over \sqrt 2}(A_{t}\sin\beta - \mu\cos\beta)\phi \equiv {h_{t}\over \sqrt 2} X_{t}\phi;
\label{mtLR2}
\end{equation}
the corresponding eigenvalues are $m_{\tilde{t}_{1}}$ and $m_{\tilde{t}_{2}}$, and the eigenstates are given by
\begin{eqnarray}
\tilde{t}_{1}& =& \cos\alpha_{t} \tilde{t}_{L} + \sin\alpha_{t} \tilde{t}_{R}\nonumber \\
\tilde{t}_{2}& =& -\sin\alpha_{t} \tilde{t}_{L} + \cos\alpha_{t} \tilde{t}_{R},
\label{mixtLtR}
\end{eqnarray}
where $\sin2\alpha_{t} = {2 m_{\tilde{t}_{LR}}^{2}\over (m_{\tilde{t}_{1}}^{2} - m_{\tilde{t}_{2}}^{2})}$. Below we will use the abbreviations $c_{t} = \cos\alpha_{t}, s_{t} = \sin\alpha_{t}$, etc.

Neglecting mixing effects in the sbottom sector,  we have 
\begin{equation}
m_{\tilde{b}_{L}}^{2} (\phi) = m_{Q}^{2} 
+ {1\over 4}(-{1\over 2} + {1\over 3} \sin^{2}\theta_{W}) (g_{w}^{2} + g'^{2})\phi^{2}\cos 2\beta
\label{mbL2}
\end{equation}

\begin{equation}
m_{\tilde{b}_{R}}^{2} (\phi) = m_{D}^{2} 
+ {1\over 4}(-{1\over 3} \sin^{2}\theta_{W}) (g_{w}^{2} + g'^{2})\phi^{2}\cos 2\beta.
\label{mbR2}
\end{equation}

For the Higgs sector, the Goldstone bosons and Higgs masses are 
\begin{eqnarray}
m_{\pi}^{2} &=& m_{H}^{2} + \lambda \phi^{2}, \nonumber \\
m_{h}^{2} &=& m_{H}^{2} + 3\lambda \phi^{2}.
\end{eqnarray}

The additional corrections that arise from supersymmetric particles can be calculated
using the two-loop unresummed potential.  Our notation
for the $D$-functions  corresponds to that of ref. \cite{KLRS}.
The contributions from the two-loop graphs containing supersymmetric particles are given below. For the $\phi$-direction,
we can drop the colour index of the squark masses:

\begin{eqnarray}
(SSV)& =& -{g_{w}^{2}\over 8} N_{c}[c_{t}^{4}D_{SSV}(m_{\tilde{t}_{1}},m_{\tilde{t}_{1}},m_{W}) + s_{t}^{4}D_{SSV}(m_{\tilde{t}_{2}},m_{\tilde{t}_{2}},m_{W})
+ 2 s_{t}^{2} c_{t}^{2}D_{SSV}(m_{\tilde{t}_{1}},m_{\tilde{t}_{2}},m_{W})
\nonumber \\
&+& D_{SSV}(m_{\tilde{b}_{L}},m_{\tilde{b}_{L}},m_{W}) +
 4 (c_{t}^{2}D_{SSV}(m_{\tilde{t}_{1}},m_{\tilde{b}_{L}},m_{W})
+ s_{t}^{2}D_{SSV}(m_{\tilde{t}_{2}},m_{\tilde{b}_{L}},m_{W})) ] \nonumber \\
&-&{g_{s}^{2}\over 4}(N_{c}^{2} -1)[D_{SSV}(m_{\tilde{t}_{1}},m_{\tilde{t}_{1}},0) +
D_{SSV}(m_{\tilde{b}_{L}},m_{\tilde{b}_{L}},0) \nonumber \\&+&
D_{SSV}(m_{\tilde{t}_{2}},m_{\tilde{t}_{2}},0)+
D_{SSV}(m_{\tilde{b}_{R}},m_{\tilde{b}_{R}},0)],
\label{SSVphi}
\end{eqnarray}

\begin{eqnarray}
(SSS)& =& -{N_{c}\over 2}\biggl\{\biggl[\biggl(h_{t}^{2} \sin^{2}\beta + {g_{w}^{2}\over 4}c_{t}^{2}\cos 2\beta\biggr)\phi + \sqrt 2 X_{t} h_{t} c_{t}s_{t}\biggr]^{2}
D_{SSS}(m_{\tilde{t}_{1}},m_{\tilde{t}_{1}},m_{h})\nonumber \\
&+&\biggl[\biggl(h_{t}^{2} \sin^{2}\beta + {g_{w}^{2}\over 4}s_{t}^{2}\cos 2\beta\biggr)\phi - \sqrt 2 X_{t} h_{t} c_{t}s_{t}\biggr]^{2}
D_{SSS}(m_{\tilde{t}_{2}},m_{\tilde{t}_{2}},m_{h})\nonumber \\
&+& 2\biggl[-\biggl( {g_{w}^{2}\over 4}c_{t} s_{t}\cos 2\beta\biggr)\phi +{1\over \sqrt 2} X_{t} h_{t} c_{2t}\biggr]^{2}
D_{SSS}(m_{\tilde{t}_{1}},m_{\tilde{t}_{2}},m_{h})\nonumber \\
&+& \biggl( {g_{w}^{2}\over 4}\cos 2\beta \phi\biggr)^{2} D_{SSS}(m_{\tilde{b}_{L}},m_{\tilde{b}_{L}},m_{h})
\nonumber \\
&+& h_{t}^{2} X_{t}^{2} D_{SSS}(m_{\tilde{t}_{1}},m_{\tilde{t}_{2}},m_{\pi}) +   \biggl[\phi \biggl(h_{t}^{2}\sin^{2}\beta+
 {g_{w}^{2}\over 2}\cos 2\beta\biggr) c_{t} + \sqrt 2 h_{t}X_{t} s_{t}\biggr]^{2}
D_{SSS}(m_{\tilde{t}_{1}},m_{\tilde{b}_{L}},m_{\pi}) \nonumber \\
&+& \biggl[\phi \biggl(h_{t}^{2}\sin^{2}\beta+
 {g_{w}^{2}\over 2}\cos 2\beta\biggr) s_{t} - \sqrt 2 h_{t}X_{t} c_{t}\biggr]^{2}
D_{SSS}(m_{\tilde{t}_{2}},m_{\tilde{b}_{L}},m_{\pi})\biggr\},
\label{SSSphi}
\end{eqnarray}

\begin{eqnarray}
(SV)& =& -{1\over 4} g_{s}^{2}(N_{c}^{2} -1)[D_{SV}(m_{\tilde{t}_{1}},0) +
 D_{SV}(m_{\tilde{b}_{L}},0) + D_{SV}(m_{\tilde{t}_{2}},0) +
D_{SV}(m_{\tilde{b}_{R}},0)]\nonumber \\
&-& {3\over 8} g_{w}^{2}N_{c}[c_{t}^{2}D_{SV}(m_{\tilde{t}_{1}},m_{W})+
s_{t}^{2} D_{SV} (m_{\tilde{t}_{2}},m_{W}) +
D_{SV}(m_{\tilde{b}_{L}},m_{W})],
\label{SVphi}
\end{eqnarray}

\begin{eqnarray}
(SS) &=& {g_{s}^{2}\over 6} N_{c}(N_{c}+1)[c_{2t}^{2}D_{SS}(m_{\tilde{t}_{1}},m_{\tilde{t}_{1}})
+ D_{SS}(m_{\tilde{b}_{R}},m_{\tilde{b}_{R}}) +
c_{2t}^{2} D_{SS}(m_{\tilde{t}_{2}},m_{\tilde{t}_{2}}) +
D_{SS}(m_{\tilde{b}_{L}},m_{\tilde{b}_{L}})] \nonumber \\
&+& {g_{s}^{2}\over 6} N_{c}^{2} (c_{t}^{4} + s_{t}^{4} + 10 c_{t}^{2} s_{t}^{2}) D_{SS}(m_{\tilde{t}_{1}}, m_{\tilde{t}_{2}}) - {g_{s}^{2} \over 2} N_{c} ( c_{t}^{4} + s_{t}^{4} - {2\over 3} s_{t}^{2} c_{t}^{2}) D_{SS}(m_{\tilde{t}_{1}}, m_{\tilde{t}_{2}}) \nonumber \\
&+& {g_{w}^{2}\over 4} N_{c}(2- N_{c})[c_{t}^{2} D_{SS}(m_{\tilde{t}_{1}},m_{\tilde{b}_{L}})+ s_{t}^{2} D_{SS}(m_{\tilde{t}_{2}}, m_{b_{L}})]\nonumber \\
&+&
h_{t}^{2}N_{c}[(c_{t}^{4} + s_{t}^{4})D_{SS}(m_{\tilde{t}_{1}},m_{\tilde{t}_{2}}) +
s_{t}^{2}D_{SS}(m_{\tilde{b}_{L}},m_{\tilde{t}_{1}})
\nonumber \\
&+& c_{t}^{2} D_{SS}(m_{\tilde{b}_{L}},m_{\tilde{t}_{2}})
+ (N_{c} + 1)c_{t}^{2} s_{t}^{2}[D_{SS}(m_{\tilde{t}_{1}},m_{\tilde{t}_{1}})
+ D_{SS}(m_{\tilde{t}_{2}},m_{\tilde{t}_{2}})] 
- 2 N_{c} s_{t}^{2} c_{t}^{2}  D_{SS}(m_{\tilde{t}_{2}},m_{\tilde{t}_{1}})] \nonumber \\
&+& \biggl({g_{w}^{2}\over 8}\biggr)N_{c}(N_{c}+
1)[ c_{t}^{4} D_{SS}(m_{\tilde{t}_{1}},m_{\tilde{t}_{1}}) +
2 c_{t}^{2} s_{t}^{2} D_{SS}(m_{\tilde{t}_{1}},m_{\tilde{t}_{2}})\nonumber \\
&+& s_{t}^{4} D_{SS}(m_{\tilde{t}_{2}},m_{\tilde{t}_{2}})+
D_{SS}(m_{\tilde{b}_{L}},m_{\tilde{b}_{L}})] \nonumber \\
&+& N_{c}\biggl[\biggl({1\over 2} h_{t}^{2}\sin^{2}\beta + {1\over 8}g_{w}^{2} \cos 2\beta c_{t}^{2}\biggr)[
D_{SS}(m_{\tilde{t}_{1}},m_{h})
+ D_{SS}(m_{\tilde{t}_{1}},m_{\pi})] \nonumber \\
&+&
\biggl({1\over 2} h_{t}^{2}\sin^{2}\beta + {1\over 8}g_{w}^{2} \cos 2\beta s_{t}^{2}\biggr)[
D_{SS}(m_{\tilde{t}_{2}},m_{h})
+ D_{SS}(m_{\tilde{t}_{2}},m_{\pi})]\biggr] \nonumber \\
&-& {1\over 8} N_{c}g_{w}^{2}\cos 2\beta[D_{SS}(m_{\tilde{b}_{L}},m_{h}) - D_{SS}(m_{\tilde{b}_{L}},m_{\pi}) + 2
c_{t}^{2} D_{SS}(m_{\tilde{t}_{1}},m_{\pi})
 + 2 s_{t}^{2}
D_{SS}(m_{\tilde{t}_{2}},m_{\pi})]\nonumber \\
&+&  N_{c} h_{t}^{2}\sin^{2}\beta[ s_{t}^{2} D_{SS}(m_{\tilde{t}_{1}},m_{\pi}) + 
c_{t}^{2} D_{SS}(m_{\tilde{t}_{2}},m_{\pi}) + D_{SS}(m_{b_{L}},m_{\pi})].
\label{SSphi}
\end{eqnarray}
There are no  additional finite contributions apart from those given in
ref. \cite{Losada3}, from counterterms
when non-zero mixing is included in  the stop sector. One can check that when all contributions are added there is no dependence on
the mixing angle in the divergent part of the potential.

\subsection{$\chi$-direction}

The gauge-boson masses are

\begin{equation}
 m_{G}^{2}= {1\over 4} g_{s}^{2} \chi^{2},
\hspace{.2in} \overline{m}_{G}^{2} = {4\over 3} m_{G}^{2}.
\label{gaugemasses}
\end{equation}
For the Higgs sector, the Goldstone bosons and Higgs masses are

\begin{eqnarray}
m_{\omega}^{2} &=& m_{\overline{\omega}}^{2} = m_{U}^{2} + \lambda_{U} \chi^{2} + h_{t}^{2}\sin^{2}\beta {\phi^{2}\over 2}, \nonumber \\
m_{u}^{2} &=& m_{U}^{2} + 3\lambda \chi^{2} + h_{t}^{2}\sin^{2}\beta {\phi^{2}\over 2}. 
\label{mpi}
\end{eqnarray}
The masses of the rest of the scalars contributing to the
effective potential are given by

\begin{equation}
m_{h_{1}}^{2} = m_{H}^{2}  + h_{t}^{2}\sin^{2}\beta {\chi^{2}\over 2},
\label{mh1}
\end{equation}

\begin{equation}
m_{\tilde{b}_{L_{1}}}^{2} = m_{Q}^{2}  + \biggl(h_{t}^{2} - {g_{s}^{2}\over 3}\biggr)
{\chi^{2}\over 2},
\label{mbL1}
\end{equation}

\begin{equation}
m_{h_{1}\tilde{b}}^{2} =- {h_{t}\over \sqrt 2} X_{t} \chi
\label{mh1b2}
\end{equation}

\begin{equation}
m_{h_{2}}^{2} = m_{H}^{2}  + h_{t}^{2}\sin^{2}\beta {\chi^{2}\over 2},
\label{mh1}
\end{equation}

\begin{equation}
m_{\tilde{t}_{L_{1}}}^{2} = m_{Q}^{2}  + \biggl(h_{t}^{2} - {g_{s}^{2}\over 3}\biggr)
{\chi^{2}\over 2},
\label{mtL1}
\end{equation}

\begin{equation}
m_{h_{2}\tilde{t}}^{2} = {h_{t}\over \sqrt 2} X_{t} \chi
\label{mh2t2}
\end{equation}
so the mixing angles are related by

\begin{equation}
\sin q_{1} = - \sin q_{2} \equiv s_{q}
\label{angles12}
\end{equation}
with eigenstates given by

\begin{eqnarray}
H_{1} &=& c_{q} h_{1} - s_{q} \tilde{b}_{L1} \nonumber \\
\tilde{q}_{2} &=& s_{q}  h_{1} + c_{q} b_{L1} \nonumber \\
H_{2} &=& c_q h_{2} + s_{q} \tilde{t}_{L1} \nonumber \\
\tilde{q}_{2} &=& - s_{q}  h_{2} + c_{q} t_{L1} 
\label{eigen2}
\end{eqnarray}

\begin{equation}
m_{\tilde{t}_{L_{2,3}}}^{2} = m_{Q}^{2}  + \biggl({g_{s}^{2}\over 6}\biggr)
{\chi^{2}\over 2},
\label{mtL2}
\end{equation}

\begin{equation}
m_{\tilde{b}_{L_{2,3}}}^{2} = m_{Q}^{2} + \biggl({g_{s}^{2}\over 6}\biggr)
{\chi^{2}\over 2},
\label{mbL2}
\end{equation}

\begin{equation}
m_{\tilde{b}_{R_{1}}}^{2} = m_{D}^{2} + \biggl({g_{s}^{2}\over 6}\biggr)
{\chi^{2}\over 2},
\label{mbR1}
\end{equation}

\begin{equation}
m_{\tilde{b}_{R_{2,3}}}^{2} = m_{D}^{2}  - \biggl({g_{s}^{2}\over 3}\biggr)
{\chi^{2}\over 2}.
\label{mbR2}
\end{equation}

In the expressions below, we also include the contributions from the
Higgs doublet.
The two-loop unresummed effective potential in the $\chi$-direction is given
by the following contributions\footnote{As $m_{\tilde{t}_{L2}} = m_{\tilde{t}_{L3}}, m_{\tilde{b}_{L2}}=m_{\tilde{b}_{L3}}, m_{\tilde{b}_{R2}}= m_{\tilde{b}_{R3}}$ in the $\chi$-direction, we just multiply 
by a factor of 2 the contributions from these fields in some of the following expressions.}:

\begin{eqnarray}
(SSV) &=& -{g_{w}^{2}\over 8}[D_{SSV}(m_{\tilde{q}_{1}},m_{\tilde{q}_{1}},0)
+ D_{SSV}(m_{H_{1}},m_{H_{1}},0) +
D_{SSV}(m_{\tilde{t}_{L2}},m_{\tilde{t}_{L2}},0) \nonumber \\ &+&
D_{SSV}(m_{\tilde{t}_{L3}},m_{\tilde{t}_{L3}},0) +
D_{SSV}(m_{\tilde{q}_{2}},m_{\tilde{q}_{2}},0)+
D_{SSV}(m_{H_{2}},m_{H_{2}},0) \nonumber \\
&+&
D_{SSV}(m_{\tilde{b}_{L2}},m_{\tilde{b}_{L2}},0) +
D_{SSV}(m_{\tilde{b}_{L3}},m_{\tilde{b}_{L3}},0)\nonumber \\
& +&
4(D_{SSV}(m_{\tilde{q}_{1}},m_{\tilde{q}_{2}},0) + D_{SSV}(m_{H_{1}}, m_{H_{2}},0)\nonumber \\
&+& D_{SSV}(m_{\tilde{t}_{L2}},m_{\tilde{b}_{L2}},0) +
D_{SSV}(m_{\tilde{t}_{L3}},m_{\tilde{b}_{L3}},0) )]\nonumber \\
&-& g_{s}^{2} {1\over 4}\biggl[(N_{c} -1) D_{SSV}(m_{\omega},\overline{m}_{\omega},m_{G})
+ (N_{c}-1)  D_{SSV}(m_{\omega},m_{u},m_{G})\nonumber \\
& +& {N_{c}-1\over N_{c}}
 D_{SSV}(\overline{m}_{\omega},m_{u},m_{G}) + {1\over N_{c}}
D_{SSV}(m_{\omega},m_{\omega},\overline{m}_{G}) 
+ N_{c}(N_{c}-2)D_{SSV}(m_{\omega},m_{\omega},0)\nonumber \\
&+& 2(N_{c} -1)(c_{q}^{2} D_{SSV}(m_{\tilde{q}_{1}},m_{\tilde{t}_{L2}},m_{G})
+ s_{q}^{2} D_{SSV}(m_{H_{2}},m_{\tilde{t}_{L2}},m_{G})) \nonumber \\
 &+& {N_{c}-1\over N_{c}} (c_{q}^{4}D_{SSV}(m_{\tilde{q}_{1}},m_{\tilde{q}_{1}},\overline{m}_{G}) + s_{q}^{4} D_{SSV}(m_{H_{2}},m_{H_{2}},\overline{m}_{G})
\nonumber \\
&+& 2 s_{q}^{2} c_{q}^{2} D_{SSV}(m_{\tilde{q}_{1}},m_{H_{2}},\overline{m}_{G}))
+ 
{1\over N_{c}} D_{SSV}(m_{\tilde{t}_{L2}},m_{\tilde{t}_{L2}},\overline{m}_{G})\nonumber \\
&+&
N_{c}(N_{c}-2) D_{SSV}(m_{\tilde{t}_{L2}},m_{\tilde{t}_{L2}},0)\nonumber \\
&+& 2(N_{c} -1) (c_{q}^{2} D_{SSV}(m_{\tilde{q}_{2}},m_{\tilde{b}_{L2}},m_{G})
+ s_{q}^{2} D_{SSV}(m_{H_{1}},m_{\tilde{b}_{L2}},m_{G}))
 \nonumber \\
&+& {N_{c}-1\over N_{c}} \biggl(c_{q}^{4}D_{SSV}(m_{\tilde{q}_{2}},m_{\tilde{q}_{2}},\overline{m}_{G}) + s_{q}^{4}D_{SSV}(m_{H_{1}},m_{H_{1}},\overline{m}_{G}) 
+ 2 s_{q}^{2} c_{q}^{2}D_{SSV}(m_{\tilde{q}_{2}},m_{H_{1}},\overline{m}_{G})\biggr)\nonumber \\
&+& 
{1\over N_{c}} D_{SSV}(m_{\tilde{b}_{L2}},m_{\tilde{b}_{L2}},\overline{m}_{G})+
N_{c}(N_{c}-2) D_{SSV}(m_{\tilde{b}_{L2}},m_{\tilde{b}_{L2}},0)\nonumber \\
&+& 2(N_{c} -1) D_{SSV}(m_{\tilde{b}_{R1}},m_{\tilde{b}_{R2}},m_{G})
 + {N_{c}-1\over N_{c}} D_{SSV}(m_{\tilde{b}_{R1}},m_{\tilde{b}_{R1}},\overline{m}_{G})\nonumber \\
& +& {1\over N_{c}} D_{SSV}(m_{\tilde{b}_{R2}},m_{\tilde{b}_{R2}},\overline{m}_{G})+
N_{c}(N_{c}-2) D_{SSV}(m_{\tilde{b}_{R2}},m_{\tilde{b}_{R2}},0)\biggr].
\label{SSVchi}
\end{eqnarray}

\begin{eqnarray}
(SV)& =& -{g_{s}^{2}\over 8}[2N_{c}(N_{c}-2) D_{SV}(m_{\omega},0) +
(N_{c} -1)[3D_{SV}(m_{\omega},m_{G}) + D_{SV}(m_{u},m_{G})] \nonumber \\
&+& {1\over N_{c}}[(N_{c} + 1)D_{SV}(m_{\omega},\overline{m}_{G})
+(N_{c}-1)D_{SV}(m_{u},\overline{m}_{G})] \nonumber \\
&+& 2N_{c}(N_{c}-2) D_{SV}(m_{\tilde{t}_{L2}},0) +
(N_{c} -1)[2D_{SV}(m_{\tilde{t}_{L2}},m_{G}) +2 (c_{q}^{2}D_{SV}(m_{\tilde{q}_{1}},m_{G}) \nonumber \\
&+& s_{q}^{2} D_{SV}(m_{H_{2}},m_{G}))]
+ {1\over N_{c}}[2D_{SV}(m_{\tilde{t}_{L2}},\overline{m}_{G})
+2(c_{q}^{2}D_{SV}(m_{\tilde{q}_{1}},\overline{m}_{G}) + s_{q}^{2} D_{SV}(m_{H_{2}},\overline{m}_{G}))] \nonumber \\
&+& 2N_{c}(N_{c}-2) D_{SV}(m_{\tilde{b}_{L2}},0) +
(N_{c} -1)[2D_{SV}(m_{\tilde{b}_{L2}},m_{G}) +2 (c_{q}^{2} D_{SV}(m_{\tilde{q}_{2}},m_{G}) \nonumber \\
&+& s_{q}^{2} D_{SV}(m_{H_{1}},m_{G}))] 
+ {1\over N_{c}}[2D_{SV}(m_{\tilde{b}_{L2}},\overline{m}_{G})
+2(c_{q}^{2}D_{SV}(m_{\tilde{q}_{2}},\overline{m}_{G}) + s_{q}^{2} D_{SV}(m_{H_{1}},\overline{m}_{G}))] \nonumber \\
&+& 2N_{c}(N_{c}-2) D_{SV}(m_{\tilde{b}_{R2}},0) +
(N_{c} -1)[2D_{SV}(m_{\tilde{b}_{R2}},m_{G}) +2 D_{SV}(m_{\tilde{b}_{R1}},m_{G})] \nonumber \\
&+& {1\over N_{c}}[2D_{SV}(m_{\tilde{b}_{R2}},\overline{m}_{G})
+2D_{SV}(m_{\tilde{b}_{R1}},\overline{m}_{G})] \nonumber \\
&-& {3\over 8} g_{w}^{2}[D_{SV}(m_{\tilde{q}_{1}},0) + D_{SV}(m_{\tilde{t}_{L2}},0) + D_{SV}(m_{\tilde{t}_{L3}},0) \nonumber \\
&+& D_{SV}(m_{\tilde{q}_{2}},0) + D_{SV}(m_{\tilde{b}_{L2}},0) + D_{SV}(m_{\tilde{b}_{L3}},0) + D_{SV}(m_{H_{1}},0) + D_{SV}(m_{H_{2}},0)],
\label{SVchi}
\end{eqnarray}

\begin{eqnarray}
(SSS) &=& -\lambda_{U}^{2} \chi^{2}[3D_{SSS}(m_{u},m_{u},m_{u}) +
(2N_{c}-1)D_{SSS}(m_{u},m_{\omega},m_{\omega})] \nonumber \\
&-& {1\over 2} \biggl(h_{t}^{2}\sin^{2}\beta c_{q}^{2}\chi +( h_{t}^{2} -{1\over 3} g_{s}^{2})s_{q}^{2}\chi + \sqrt 2 X_{t} h_{t} c_{q} s_{q}\biggr)^{2}[D_{SSS}(m_{u},m_{H_{1}},m_{H_{1}})\nonumber \\
&+&
D_{SSS}(m_{u}, m_{H_{2}},m_{H_{2}})] \nonumber \\
&-& {1\over 2}\biggl(h_{t}^{2} \sin^{2}\beta s_{q}^{2}\chi +(h_{t}^{2} -{1\over 3} g_{s}^{2})c_{q}^{2}\chi - \sqrt 2 X_{t} h_{t} c_{q} s_{q}\biggr)^{2}
[D_{SSS}(m_{u},m_{\tilde{q}_{1}},m_{\tilde{q}_{1}}) \nonumber \\
&+&
D_{SSS}(m_{u},m_{\tilde{q}_{2}},m_{\tilde{q}_{2}})]\nonumber \\
&-&
2\biggl((h_{t}^{2} -{1\over 2} g_{s}^{2}){\chi \over \sqrt2}c_{q} - s_{q} X_{t} h_{t}\biggr)^{2} [D_{SSS}(m_{\omega},m_{\tilde{q}_{1}},m_{\tilde{t}_{L2}}) 
+ D_{SSS}(m_{\omega},m_{\tilde{q}_{2}},m_{\tilde{b}_{L2}}]\nonumber \\
&-&
2[(h_{t}^{2} -{1\over 2} g_{s}^{2}){\chi \over \sqrt2}s_{q} + c_{q} X_{t} h_{t}]^{2} [D_{SSS}(m_{\omega},m_{H_{2}},m_{\tilde{t}_{L2}})) 
+D_{SSS}(m_{\omega},m_{H_{1}},m_{\tilde{b}_{L2}})]
 \nonumber \\
&-& (-h_{t}^{2}\sin^{2}\beta s_{q}c_{q} \chi + c_{2q}{X_{t}\over \sqrt 2} + s_{q} c_{q} (h_{t}^{2} -{1\over 3} g_{s}^{2})\chi)^{2} D_{SSS}(m_{u},m_{H_{2}}, m_{q_{1}})
\nonumber \\
&-& (h_{t}^{2}\sin^{2}\beta s_{q}c_{q} \chi + c_{2q} {X_{t}\over \sqrt 2} -  s_{q} c_{q} (h_{t}^{2} -{1\over 3} g_{s}^{2})\chi)^{2} D_{SSS}(m_{u},m_{H_{1}}, m_{q_{2}})
\nonumber \\
&-& [c_{2q} {X_{t}\over \sqrt 2} h_{t}]^{2}(D_{SSS}(m_{\omega},m_{H_{1}}, m_{q_{2}}) + D_{SSS}(m_{\omega},m_{H_{2}}, m_{q_{1}}))\nonumber \\
&-& {\chi^{2}\over 2}\biggl( 2({g_{s}^{2}\over 6})^{2}D_{SSS}(m_{u},m_{\tilde{t}_{L2}},m_{\tilde{t}_{L2}})+ 2({g_{s}^{2}\over 6})^{2}D_{SSS}(m_{u},m_{\tilde{b}_{L2}},m_{\tilde{b}_{L2}}) \nonumber \\
&+&({1\over 3} g_{s}^{2})^{2}
D_{SSS}(m_{u},m_{\tilde{b}_{R1}},m_{\tilde{b}_{R1}}) +
2({1\over 2} g_{s}^{2})^{2}D_{SSS}(m_{\omega},m_{\tilde{b}_{R1}},m_{\tilde{b}_{R2}})\nonumber \\
&+& 2({g_{s}^{2}\over 6})^{2}D_{SSS}(m_{u},m_{\tilde{b}_{R2}},m_{\tilde{b}_{R2}})\biggr) ,
\label{SSSchi}
\end{eqnarray}

\begin{eqnarray}
(SS) &=& 2\lambda \biggl[c_{q}^{4} (D_{SS}(m_{h_{1}},m_{h_{1}}) +
D_{SS}(m_{h_{2}},m_{h_{2}}) + D_{SS}(m_{h_{1}},m_{h_{2}})) \nonumber \\
&+&  s_{q}^{4}
 (D_{SS}(m_{\tilde{q}_{1}},m_{\tilde{q}_{1}}) +
D_{SS}(m_{\tilde{q}_{2}},m_{\tilde{q}_{2}}) + D_{SS}(m_{\tilde{q}_{1}},m_{\tilde{q}_{2}})) \nonumber \\
&+& c_{q}^{2} s_{q}^{2}( 2 D_{SS}(m_{h_{1}},m_{\tilde{q}_{2}}) +
2 D_{SS}(m_{h_{2}},m_{\tilde{q}_{1}}) + D_{SS}(m_{h_{1}},m_{\tilde{q}_{1}})
+ D_{SS}(m_{h_{2}}, m_{\tilde{q}_{2}}))\biggr] \nonumber \\
&+& \biggl({g_{w}^{2}\over 8} + {g_{s}^{2}\over 6}\biggr)[2 c_{q}^{4} D_{SS}(m_{\tilde{q}_{1}},m_{\tilde{q}_{1}}) +
2 s_{q}^{4} D_{SS}(m_{h_{2}},m_{h_{2}}) +
4 s_{q}^{2} c_{q}^{2} D_{SS}(m_{h_{2}},m_{\tilde{q}_{1}}) \nonumber \\
&+& 4(s_{q}^{2} D_{SS}(m_{h_{2}},m_{\tilde{t}_{L2}}) + 
c_{q}^{2} D_{SS}(m_{\tilde{q}_{1}},m_{\tilde{t}_{L2}})) +
6 D_{SS}(m_{\tilde{t}_{L2}},m_{\tilde{t}_{L2}}) \nonumber \\
&+&2 c_{q}^{4} D_{SS}(m_{\tilde{q}_{2}},m_{\tilde{q}_{2}}) +
2 s_{q}^{4} D_{SS}(m_{h_{1}},m_{h_{1}}) +
4 s_{q}^{2} c_{q}^{2} D_{SS}(m_{h_{1}},m_{\tilde{q}_{2}}) \nonumber \\
&+& 4(s_{q}^{2} D_{SS}(m_{h_{1}},m_{\tilde{t}_{L2}}) + 
c_{q}^{2} D_{SS}(m_{\tilde{q}_{2}},m_{\tilde{t}_{L2}})) +
6 D_{SS}(m_{\tilde{b}_{L2}},m_{\tilde{b}_{L2}})] \nonumber \\
&+& {1\over 4}( {g_{s}^{2}\over 6})[3 D_{SS}(m_{u},m_{u}) +
10 D_{SS}(m_{u},m_{\omega}) + 35 D_{SS}(m_{\omega},m_{\omega})\nonumber \\
&+& 8 D_{SS}(m_{\tilde{b}_{R1}},m_{\tilde{b}_{R1}}) +
24 D_{SS}(m_{\tilde{b}_{R2}},m_{\tilde{b}_{R2}}) + 16 D_{SS}(m_{\tilde{b}_{R1}},m_{\tilde{b}_{R2}})]\nonumber \\
&+& {1\over 2} (h_{t}^{2}- {1\over 2} g_{s}^{2})[c_{q}^{2} D_{SS}(m_{\tilde{q}_{1}},m_{u}) + c_{q}^{2} D_{SS}(m_{\tilde{q}_{1}},m_{\omega})
+ s_{q}^{2} D_{SS}(m_{h_{2}},m_{u}) + s_{q}^{2} D_{SS}(m_{h_{2}},m_{\omega})\nonumber \\
& +& 2 D_{SS}(m_{\tilde{t}_{L2}},m_{\omega})
+ 2 D_{SS}(m_{\tilde{t}_{L3}},m_{\omega})] \nonumber \\ 
&+& {1\over 12} g_{s}^{2}[c_{q}^{2}D_{SS}(m_{u},m_{\tilde{q}_{1}})
+ 5 c_{q}^{2} D_{SS}(m_{\omega},m_{\tilde{q}_{1}}) +
s_{q}^{2}D_{SS}(m_{u},m_{h_{2}})\nonumber \\
&+& 5 s_{q}^{2} D_{SS}(m_{\omega},m_{h_{2}}) +
D_{SS}(m_{u},m_{\tilde{t}_{L2}}) \nonumber \\
&+& 5 D_{SS}(m_{\omega},m_{\tilde{t}_{L2}}) +
D_{SS}(m_{u},m_{\tilde{t}_{L3}})
+ 5 D_{SS}(m_{\omega},m_{\tilde{t}_{L3}})]\nonumber \\
&+& {1\over 2} (h_{t}^{2}- {1\over 2} g_{s}^{2})[c_{q}^{2} D_{SS}(m_{\tilde{q}_{2}},m_{u}) + c_{q}^{2} D_{SS}(m_{\tilde{q}_{2}},m_{\omega})\nonumber \\
&+& s_{q}^{2} D_{SS}(m_{h_{1}},m_{u}) + s_{q}^{2} D_{SS}(m_{h_{1}},m_{\omega}) + 2 D_{SS}(m_{\tilde{t}_{L2}},m_{\omega})
+ 2 D_{SS}(m_{\tilde{t}_{L3}},m_{\omega})] \nonumber \\ 
&+& {1\over 12} g_{s}^{2}[c_{q}^{2}D_{SS}(m_{u},m_{\tilde{q}_{2}})
+ 5 c_{q}^{2} D_{SS}(m_{\omega},m_{\tilde{q}_{2}}) +
s_{q}^{2}D_{SS}(m_{u},m_{h_{1}})\nonumber \\
&+& 5 s_{q}^{2} D_{SS}(m_{\omega},m_{h_{1}}) +
D_{SS}(m_{u},m_{\tilde{b}_{L2}}) \nonumber \\
&+& 5 D_{SS}(m_{\omega},m_{\tilde{b}_{L2}}) +
D_{SS}(m_{u},m_{\tilde{b}_{L3}})
+ 5 D_{SS}(m_{\omega},m_{\tilde{b}_{L3}})]\nonumber \\
&+& \biggl( h_{t}^{2}\sin^{2}\beta + {1\over 4} g_{w}^{2}\cos 2\beta\biggr) \biggl[2 c_{q}^{2} s_{q}^{2}[D_{SS}(m_{h_{1}},m_{h_{1}}) \nonumber \\
&+& D_{SS}(m_{h_{2}},m_{h_{2}})
+  D_{SS}(m_{h_{1}},m_{h_{2}}) +  D_{SS}(m_{\tilde{q}_{1}},m_{\tilde{q}_{2}})
+ D_{SS}(m_{\tilde{q}_{1}},m_{\tilde{q}_{1}})\nonumber \\
& +&
D_{SS}(m_{\tilde{q}_{2}},m_{\tilde{q}_{2}})]
+ c_{q}^{2}[D_{SS}(m_{\tilde{t}_{L2}},m_{h_{2}})  + D_{SS}(m_{\tilde{t}_{L3}},m_{h_{2}}) \nonumber \\
&+&  D_{SS}(m_{\tilde{b}_{L2}},m_{h_{1}})  + D_{SS}(m_{\tilde{b}_{L3}},m_{h_{1}})] + s_{q}^{2} [D_{SS}(m_{\tilde{t}_{L2}},m_{\tilde{q}_{1}})  + D_{SS}(m_{\tilde{t}_{L3}},m_{\tilde{q}_{1}}) \nonumber \\
&+&  D_{SS}(m_{\tilde{b}_{L2}},m_{\tilde{q}_{2}})  + D_{SS}(m_{\tilde{b}_{L3}},m_{\tilde{q}_{2}})]+
 c_{2q}^{2}[D_{SS}(m_{\tilde{q}_{1}},m_{h_{2}})
+ D_{SS}(m_{\tilde{q}_{2}},m_{h_{1}})] \biggr]\nonumber \\
&+& \biggl({1\over 4} g_{w}^{2}(c_{q}^{4} + s_{q}^{4}) -  (c_{q}^{4} + s_{q}^{4}) {1\over 2}
g_{w}^{2} \cos^{2} \beta - 2 s_{q}^{2} c_{q}^{2}(-{1\over2} g_{w}^{2}\sin^{2} \beta + h_{t}^{2} \sin^{2}\beta)\biggr)[D_{SS}(m_{\tilde{q}_{1}},m_{h_{1}})\nonumber \\
& +& D_{SS}(m_{\tilde{q}_{2}},m_{h_{2}})] - {g_{w}^{2} \over 2} s_{q}^{2} c_{q}^{2}[ D_{SS}(m_{h_{1}}, m_{h_{2}}) + D_{SS} (m_{\tilde{q}_{1}},m_{\tilde{q}_{2}})] \nonumber \\
&-&  \biggl( {1\over 4} g_{w}^{2}\cos 2\beta\biggr)[ c_{q}^{2}(D_{SS}(m_{\tilde{b}_{L2}},m_{h_{2}})  + D_{SS}(m_{\tilde{b}_{L3}},m_{h_{2}}) \nonumber \\
&+&  D_{SS}(m_{\tilde{t}_{L2}},m_{h_{1}})  + D_{SS}(m_{\tilde{t}_{L3}},m_{h_{1}})) + s_{q}^{2}(D_{SS}(m_{\tilde{b}_{L2}},m_{\tilde{q}_{1}})  + D_{SS}(m_{\tilde{b}_{L3}},m_{\tilde{q}_{1}}) \nonumber \\
&+&  D_{SS}(m_{\tilde{t}_{L2}},m_{\tilde{q}_{2}})  + D_{SS}(m_{\tilde{t}_{L3}},m_{\tilde{q}_{2}})) ]\nonumber \\
&+& {1\over 2} h_{t}^{2}\sin^{2}\beta[c_{q}^{2} D_{SS}(m_{u},m_{h_{1}})
+ c_{q}^{2} D_{SS}(m_{u},m_{h_{2}}) + s_{q}^{2} D_{SS}(m_{u},m_{\tilde{q}_{1}})
+ s_{q}^{2} D_{SS}(m_{u},m_{\tilde{q}_{2}})\nonumber \\
&+& 5 (c_{q}^{2} D_{SS}(m_{\omega},m_{h_{1}})
+ c_{q}^{2} D_{SS}(m_{\omega},m_{h_{2}}) + s_{q}^{2} D_{SS}(m_{\omega},m_{\tilde{q}_{1}})
+ s_{q}^{2} D_{SS}(m_{\omega},m_{\tilde{q}_{2}}))]  \nonumber \\
&+& {g_{w}^{2}\over 4} (2-N_{c})[c_{q}^{4} D_{SS}(m_{\tilde{q}_{1}},m_{\tilde{q}_{2}})\nonumber \\
& +& s_{q}^{4} D_{SS}(m_{h_{1}},m_{h_{2}}) + s_{q}^{2} c_{q}^{2} D_{SS}(m_{h_{2}},m_{\tilde{q}_{2}}) + s_{q}^{2} c_{q}^{2} D_{SS}(m_{h_{1}}, m_{\tilde{q}_{1}}) \nonumber \\
&+&  D_{SS}(m_{\tilde{t}_{L2}},m_{\tilde{b}_{L2}}) + D_{SS}(m_{\tilde{t}_{L3}},m_{\tilde{b}_{L3}})] \nonumber \\
&+&  {1\over 2}g_{s}^{2}[ c_{q}^{4}D_{SS}(m_{\tilde{q}_{1}},m_{\tilde{q}_{2}}) +
s_{q}^{4} D_{SS}(m_{h_{1}}, m_{h_{2}}) + s_{q}^{2} c_{q}^{2} (D_{SS}(m_{\tilde{q}_{1}}, m_{h_{1}}) \nonumber \\
&+&D_{SS}(m_{\tilde{q}_{2}}, m_{h_{2}})) +
D_{SS}(m_{\tilde{t}_{L2}}, m_{\tilde{b}_{L2}})
+ D_{SS}(m_{\tilde{b}_{L3}},m_{\tilde{t}_{L3}})]\nonumber \\
&-& {1\over 6}g_{s}^{2}[ c_{q}^{4}D_{SS}(m_{\tilde{q}_{1}},m_{\tilde{q}_{2}}) +
s_{q}^{4} D_{SS}(m_{h_{1}}, m_{h_{2}}) + s_{q}^{2} c_{q}^{2} (D_{SS}(m_{\tilde{q}_{1}}, m_{h_{1}}) \nonumber \\
&+& D_{SS}(m_{\tilde{q}_{2}}, m_{h_{2}}))+
c_{q}^{2}D_{SS}(m_{\tilde{q}_{1}},m_{\tilde{b}_{L2}})+ s_{q}^{2} D_{SS}(m_{h_{2}}, m_{\tilde{b}_{L2}}) \nonumber \\
&+& c_{q}^{2} D_{SS}(m_{\tilde{q}_{1}},m_{\tilde{b}_{L3}}) + s_{q}^{2} D_{SS}(m_{h_{2}}, m_{\tilde{b}_{L3}})\nonumber \\
&+& c_{q}^{2} D_{SS}(m_{\tilde{t}_{L2}},m_{\tilde{q}_{2}}) + s_{q}^{2} D_{SS}(m_{\tilde{t}_{L2}}, m_{h_{1}}) + D_{SS}(m_{\tilde{t}_{L2}},m_{\tilde{b}_{L2}})+
D_{SS}(m_{\tilde{t}_{L2}},m_{\tilde{b}_{L3}})\nonumber \\
& +& c_{q}^{2}D_{SS}(m_{\tilde{t}_{L3}},m_{\tilde{q}_{2}}) + s_{q}^{2} D_{SS}(m_{\tilde{t}_{L3}}, m_{h_{1}})
+ D_{SS}(m_{\tilde{t}_{L3}},m_{\tilde{b}_{L2}})\nonumber \\
&+& D_{SS}(m_{\tilde{t}_{L3}},m_{\tilde{b}_{L3}})\nonumber \\
&-& {1\over 2} g_{s}^{2}[c_{q}^{2} D_{SS}(m_{\tilde{q}_{1}},m_{\tilde{b}_{R1}})
+ s_{q}^{2} D_{SS}(m_{h_{2}}, m_{\tilde{b}_{R1}}) + D_{SS}(m_{\tilde{t}_{L2}},m_{\tilde{b}_{R2}})\nonumber \\
&+& D_{SS}(m_{\tilde{t}_{L3}},m_{\tilde{b}_{R3}}) + c_{q}^{2} D_{SS}(m_{\tilde{q}_{2}},m_{\tilde{b}_{R1}}) + s_{q}^{2} D_{SS}(m_{h_{1}}, m_{\tilde{b}_{R1}}) \nonumber \\
&+& D_{SS}(m_{\tilde{b}_{L2}},m_{\tilde{b}_{R2}})
+ D_{SS}(m_{\tilde{b}_{L3}},m_{\tilde{b}_{R3}})]\nonumber \\
&+& {1\over 6}g_{s}^{2}(c_{q}^{2} D_{SS}(m_{\tilde{q}_{1}},m_{\tilde{b}_{R1}})
+ s_{q}^{2} D_{SS}(m_{h_{2}}, m_{\tilde{b}_{R1}}) \nonumber \\
&+&c_{q}^{2} D_{SS}(m_{\tilde{q}_{1}},m_{\tilde{b}_{R2}})
+ s_{q}^{2} D_{SS}(m_{h_{2}}, m_{\tilde{b}_{R2}})  + c_{q}^{2} D_{SS}(m_{\tilde{q}_{1}},m_{\tilde{b}_{R3}})
+ s_{q}^{2} D_{SS}(m_{h_{2}}, m_{\tilde{b}_{R3}}) \nonumber \\ 
&+& D_{SS}(m_{\tilde{t}_{L2}},m_{\tilde{b}_{R1}}) + D_{SS}(m_{\tilde{t}_{L2}},m_{\tilde{b}_{R2}})+
D_{SS}(m_{\tilde{t}_{L2}},m_{\tilde{b}_{R3}})\nonumber \\
& +& D_{SS}(m_{\tilde{t}_{L3}},m_{\tilde{b}_{R1}})
+ D_{SS}(m_{\tilde{t}_{L3}},m_{\tilde{b}_{R2}})\nonumber \\
&+& D_{SS}(m_{\tilde{t}_{L3}},m_{\tilde{b}_{R3}}) 
\nonumber \\
&+&c_{q}^{2} D_{SS}(m_{\tilde{q}_{2}},m_{\tilde{b}_{R1}})
+ s_{q}^{2} D_{SS}(m_{h_{1}}, m_{\tilde{b}_{R1}}) \nonumber \\
&+&c_{q}^{2} D_{SS}(m_{\tilde{q}_{2}},m_{\tilde{b}_{R2}})
+ s_{q}^{2} D_{SS}(m_{h_{1}}, m_{\tilde{b}_{R2}})  + c_{q}^{2} D_{SS}(m_{\tilde{q}_{2}},m_{\tilde{b}_{R3}})
+ s_{q}^{2} D_{SS}(m_{h_{1}}, m_{\tilde{b}_{R3}}) \nonumber \\
&+& D_{SS}(m_{\tilde{b}_{L2}},m_{\tilde{b}_{R1}}) + D_{SS}(m_{\tilde{b}_{L2}},m_{\tilde{b}_{R2}})+
D_{SS}(m_{\tilde{b}_{L2}},m_{\tilde{b}_{R3}}) + D_{SS}(m_{\tilde{b}_{L3}},m_{\tilde{b}_{R1}})\nonumber \\
&+& D_{SS}(m_{\tilde{b}_{L3}},m_{\tilde{b}_{R2}})
+ D_{SS}(m_{\tilde{b}_{L3}},m_{\tilde{b}_{R3}})]\nonumber \\
&+& {1\over 4} g_{s}^{2}[D_{SS}(m_{u},m_{\tilde{b}_{R1}}) + D_{SS}(m_{\omega},m_{\tilde{b}_{R1}}) + 2 D_{SS}(m_{\omega},m_{\tilde{b}_{R2}})
+ 2 D_{SS}(m_{\omega},m_{\tilde{b}_{R3}})]\nonumber \\
&-& {1\over 12} g_{s}^{2}[D_{SS}(m_{u},m_{\tilde{b}_{R1}})+
D_{SS}(m_{u},m_{\tilde{b}_{R2}}) + D_{SS}(m_{u},m_{\tilde{b}_{R3}})\nonumber \\
&+& 5 D_{SS}(m_{\omega},m_{\tilde{b}_{R1}}) +5 D_{SS}(m_{\omega},m_{\tilde{b}_{R2}})+
5D_{SS}(m_{\omega},m_{\tilde{b}_{R3}})].
\label{SSchimix}
\end{eqnarray}

\subsection{Integration over the heavy scale}

The second part of the calculation arises, as noticed in the paper by Kajantie et al. \cite{KLRS}: when  the ``heavy" particles have been integrated out
 their contributions
to the 3D mass parameters  should also be included, as they can substantially
vary the value of the parameters $\Lambda_{H_{3}}, \Lambda_{U_{3}}$. In order to do this we must
calculate the two-loop contributions to the effective potential in the $\phi$-
and $\chi$-directions from the heavy fields: $Q, D, C_{o}, A_{o}$. In the following rotation to eigenstates the angle is temperature-dependent; however, if we verify that the eigenstates are always
well separated close to the transition point, there is no
ambiguity about which is the field that is being integrated out
at the second stage.

The masses in the shifted theory are now given by

\begin{equation}
m_{\tilde{t}_{L_{1}}}^{2} = \overline{m}_{Q_{3}}^{2} + (h_{t}^{L} + \Lambda_{3} + 
\Lambda_{4}^{s}){\phi^{2}\over 2} + (h_{t}^{QU} + g_{s_{1}}^{QU} + g_{s_{2}}^{QU})
{\chi^{2}\over 2},
\label{mmtL1}
\end{equation}

\begin{equation}
m_{\tilde{t}_{L_{2,3}}}^{2} = \overline{m}_{Q_{3}}^{2} + (h_{t}^{L} + \Lambda_{3} + 
\Lambda_{4}^{s}){\phi^{2}\over 2} + (g_{s_{2}}^{QU})
{\chi^{2}\over 2},
\label{mmtL2}
\end{equation}

\begin{equation}
m_{\tilde{b}_{L_{1}}}^{2} = \overline{m}_{Q_{3}}^{2} + ( \Lambda_{3} + 
\Lambda_{4}^{c}){\phi^{2}\over 2} + (h_{t}^{QU} + g_{s_{1}}^{QU} + g_{s_{2}}^{QU})
{\chi^{2}\over 2},
\label{mmbL1}
\end{equation}

\begin{equation}
m_{\tilde{b}_{L_{2,3}}}^{2} = \overline{m}_{Q_{3}}^{2} + (\Lambda_{3} + 
\Lambda_{4}^{c}){\phi^{2}\over 2} + (g_{s_{2}}^{QU})
{\chi^{2}\over 2},
\label{mmbL2}
\end{equation}

\begin{equation}
m_{\tilde{b}_{R_{1}}}^{2} = m_{D_{3}}^{2} + (g_{s_{1}}^{UD} + g_{s_{2}}^{UD})
{\chi^{2}\over 2},
\label{mmbR1}
\end{equation}

\begin{equation}
m_{\tilde{b}_{R_{2,3}}}^{2} = \overline{m}_{D_{3}}^{2}  + (g_{s_{2}}^{QU})
{\chi^{2}\over 2}.
\label{mmbR2}
\end{equation}
and the relevant mixing terms are:
\begin{equation}
m_{\tilde{t}_{LR}}^{2} (\phi)= {h_{t}\over \sqrt 2}(\overline{A}_{t}\sin\beta - \overline{\mu}\cos\beta)\phi \equiv {h_{t}\over \sqrt 2} \overline{X}_{t}\phi
\label{mtLR2nd}
\end{equation}
for the $\phi$-direction, and

\begin{equation}
m_{\tilde{t}_{\pm}}^{2} (\phi)= \pm   {h_{t}\over \sqrt 2} \overline{X}_{t}\chi
\label{mtqh12}
\end{equation}
for the $\chi$-direction.

The expressions for the rest of the fields are given in \cite{Laine2}.
The two-loop contributions from the heavy scale are given below. We stress
that the $D$-integrals in eqs. (\ref{Vheavyphi}) and (\ref{V2heavy}) are just $3D$ integrals, our notation
follows that of refs. \cite{Laine2, farakos, KLRS}\footnote{Our convention
for the functions $ D_{VVV}, D_{VVS}$ is that of \cite{Laine2}.}.
We do not write the contributions arising from the longitudinal components
of the gauge fields $A_{o}$ and $C_{o}$, since the only modification that
is necessary is to substitute the gauge couplings $g_{w(s)} \rightarrow g_{w(s)_{3}}$, 
which are given in ref. \cite{Losada3} .

\subsubsection{$\phi$-direction}

\begin{eqnarray}
V_{2}^{heavy}(\phi) & =&- {g_{w_{3}}^{2}\over 8} N_{c}[c_{t}^{4}D_{SSV}(m_{\tilde{t}_{1}},m_{\tilde{t}_{1}},m_{W}) + s_{t}^{4}D_{SSV}(m_{\tilde{t}_{2}},m_{\tilde{t}_{2}},m_{W})
+ 2 s_{t}^{2} c_{t}^{2}D_{SSV}(m_{\tilde{t}_{1}},m_{\tilde{t}_{2}},m_{W})
\nonumber \\
&+& D_{SSV}(m_{\tilde{b}_{L}},m_{\tilde{b}_{L}},m_{W}) +
 4 (c_{t}^{2}D_{SSV}(m_{\tilde{t}_{1}},m_{\tilde{b}_{L}},m_{W})
+ s_{t}^{2}D_{SSV}(m_{\tilde{t}_{2}},m_{\tilde{b}_{L}},m_{W}))] \nonumber \\
&-&{g_{s_{3}}^{2}\over 4}(N_{c}^{2} -1)[c_{t}^{4}D_{SSV}(m_{\tilde{t}_{1}},m_{\tilde{t}_{1}},0) + s_{t}^{4} D_{SSV}(m_{\tilde{t}_{2}}, m_{\tilde{t}_{2}},0) +
2 c_{t}^{2}s_{t}^{2} D_{SSV}(m_{\tilde{t}_{1}}, m_{\tilde{t}_{2}},0) \nonumber \\
&+&
D_{SSV}(m_{\tilde{b}_{L}},m_{\tilde{b}_{L}},0) 
+
D_{SSV}(m_{\tilde{b}_{R}},m_{\tilde{b}_{R}},0)]\nonumber \\
& -&{N_{c}\over 2}\biggl\{\biggl[\biggl((h_{t}^{L}  + \Lambda_{3} + \Lambda_{4}^{s})c_{t}^{2}\biggr)\phi + \sqrt 2 \overline{X}_{t} h_{t} c_{t}s_{t}\biggr]^{2}
D_{SSS}(m_{\tilde{t}_{1}},m_{\tilde{t}_{1}},m_{h})\nonumber \\
&+&\biggl[(h_{t}^{L}  + \Lambda_{3} + \Lambda_{4}^{s})s_{t}^{2}\biggr)\phi - \sqrt 2 X_{t} h_{t} c_{t}s_{t}\biggr]^{2}
D_{SSS}(m_{\tilde{t}_{2}},m_{\tilde{t}_{2}},m_{h})\nonumber \\
&+& 2\biggl[-\biggl( (h_{t}^{L}  + \Lambda_{3} + \Lambda_{4}^{s})c_{t} s_{t}
\biggr)\phi +{1\over \sqrt 2} \overline{X}_{t} h_{t} c_{2t}\biggr]^{2}
D_{SSS}(m_{\tilde{t}_{1}},m_{\tilde{t}_{2}},m_{h})\nonumber \\
&+& \biggl( \Lambda_{3} + \Lambda_{4}^{c}\biggr)^{2}\phi D_{SSS}(m_{\tilde{b}_{L}},m_{\tilde{b}_{L}},m_{h})
\nonumber \\
&+& h_{t}^{2} \overline{X}_{t}^{2} D_{SSS}(m_{\tilde{t}_{1}},m_{\tilde{t}_{2}},m_{\pi}) +   \biggl[\phi \biggl(h_{t}^{L}-\Lambda_{4}^{c} + \Lambda_4^{s}\biggr) c_{t} + \sqrt 2 h_{t}\overline{X}_{t} s_{t}\biggr]^{2}
D_{SSS}(m_{\tilde{t}_{1}},m_{\tilde{b}_{L}},m_{\pi}) \nonumber \\
&+& \biggl[\phi \biggl(h_{t}^{L}-\Lambda_{4}^{c} + \Lambda_4^{s}\biggr) s_{t} - \sqrt 2 h_{t}\overline{X}_{t} c_{t}\biggr]^{2}
D_{SSS}(m_{\tilde{t}_{2}},m_{\tilde{b}_{L}},m_{\pi})\biggr\} \nonumber \\
& -& {1\over 4} g_{s_{3}}^{2}(N_{c}^{2} -1)[c_{t}^{2} D_{SV}(m_{\tilde{t}_{1}},0) + s_{t}^{2} D_{SV}(m_{\tilde{t}_{2}},0)
+ D_{SV}(m_{\tilde{b}_{L}},0)  +
D_{SV}(m_{\tilde{b}_{R}},0)]\nonumber \\
&-& {3\over 8} g_{w_{3}}^{2}N_{c}[ c_{t}^{2} D_{SV}(m_{\tilde{t}_{1}},m_{W})+
+ s_{t}^{2} D_{SV}(m_{\tilde{t}_{2}},m_{W})+
D_{SV}(m_{\tilde{b}_{L}},m_{W})]\nonumber \\
&+& N_{c}(N_{c}+1)[\lambda_{Q_{3}}c_{t}^{4} + g_{s_{1}}^{QU} s_{t}^{2} c_{t}^{2} + g_{s_{2}}^{QU} c_{t}^{2} s_{t}^{2}] D_{SS}(m_{\tilde{t}_{1}},m_{\tilde{t}_{1}})
\nonumber \\
& +&  N_{c}(N_{c}+1)[\lambda_{Q_{3}}s_{t}^{4} + g_{s_{1}}^{QU} s_{t}^{2} c_{t}^{2} + g_{s_{2}}^{QU} c_{t}^{2} s_{t}^{2}] D_{SS}(m_{\tilde{t}_{2}},m_{\tilde{t}_{2}}) +
 \lambda_{Q_{3}} N_{c} (N_{c} +1)D_{SS}(m_{\tilde{b}_{L}},m_{\tilde{b}_{L}}) \nonumber \\
 &+& N_{c}^{2} (2 c_{t}^{2} s_{t}^{2} \lambda_{Q_{3}} - 2 c_{t}^{2} s_{t}^{2} g_{s_{1}}^{QU} + (c_{t}^{4} + s_{t}^{4}) g_{s_{2}}^{QU}) D_{SS} (m_{\tilde{t}_{1}}, m_{\tilde{t}_{2}}) \nonumber \\
&+& N_{c} ( 2 c_{t}^{2} s_{t}^{2} \lambda_{Q_{3}} - 2 c_{t}^{2} s_{t}^{2} g_{s_{2}}^{QU} + (c_{t}^{4} + s_{t}^{4}) g_{s_{1}}^{QU}) D_{SS} (m_{\tilde{t}_{1}}, m_{\tilde{t}_{2}}) \nonumber \\
&+&\lambda_{D_{3}} N_{c}(N_{c}+1) D_{SS}(m_{\tilde{b}_{R}},m_{\tilde{b}_{R}})+
  (2\Lambda_{1} ) N_{c}(2- N_{c})[c_{t}^{2} D_{SS}(m_{\tilde{t}_{1}},m_{\tilde{b}_{L}}) \nonumber \\
&+& s_{t}^{2} D_{SS}(m_{\tilde{t}_{2}}, m_{b_{L}})]\nonumber \\
& +& ( N_{c}g_{s_{1}}^{QQ} + N_{c}^{2}g_{s_{2}}^{QQ})[c_{t}^{2} D_{SS}(m_{\tilde{t}_{1}},m_{\tilde{b}_{L}}) \nonumber \\
&+& s_{t}^{2} D_{SS}(m_{\tilde{t}_{2}}, m_{b_{L}})]\nonumber \\
&+& N_{c} (g_{s_{1}}^{QU} + N_{c} g_{s_{2}}^{QU})[s_{t}^{2} D_{SS}(m_{\tilde{t}_{1}}, m_{\tilde{b}_{L}}) + c_{t}^{2} D_{SS}(m_{\tilde{t}_{2}}, m_{\tilde{b}_{L}})] \nonumber \\
&+&
h_{t}^{QU}  N_{c}[(c_{t}^{4} + s_{t}^{4})D_{SS}(m_{\tilde{t}_{1}},m_{\tilde{t}_{2}}) +
s_{t}^{2}D_{SS}(m_{\tilde{b}_{L}},m_{\tilde{t}_{1}})
\nonumber \\
&+& c_{t}^{2} D_{SS}(m_{\tilde{b}_{L}},m_{\tilde{t}_{2}})
+ (N_{c} + 1)c_{t}^{2} s_{t}^{2}[D_{SS}(m_{\tilde{t}_{1}},m_{\tilde{t}_{1}})
+ D_{SS}(m_{\tilde{t}_{2}},m_{\tilde{t}_{2}})] 
- 2  D_{SS}(m_{\tilde{t}_{2}},m_{\tilde{t}_{1}})] \nonumber \\
&+& \Lambda_{1}N_{c}(N_{c}+
1)[ c_{t}^{4} D_{SS}(m_{\tilde{t}_{1}},m_{\tilde{t}_{1}}) +
2 c_{t}^{2} s_{t}^{2} D_{SS}(m_{\tilde{t}_{1}},m_{\tilde{t}_{2}}) \nonumber \\
&+& s_{t}^{4} D_{SS}(m_{\tilde{t}_{2}},m_{\tilde{t}_{2}})+
D_{SS}(m_{\tilde{b}_{L}},m_{\tilde{b}_{L}})] \nonumber \\
&+& {N_{c}\over 2}\biggl[\biggl( (h_{t}^{L} + \Lambda_{3} + \Lambda_{4}^{s}) c_{t}^{2}\biggr)[
D_{SS}(m_{\tilde{t}_{1}},m_{h})
+ D_{SS}(m_{\tilde{t}_{1}},m_{\pi})] \nonumber \\
&+&
\biggl( (h_{t}^{L} + \Lambda_{3} + \Lambda_{4}^{s}) s_{t}^{2}\biggr)[
D_{SS}(m_{\tilde{t}_{2}},m_{h})
+ D_{SS}(m_{\tilde{t}_{2}},m_{\pi})]\biggr] \nonumber \\
&-& {1\over 2} N_{c}(\Lambda_{3} + \Lambda_{4}^{c})[D_{SS}(m_{\tilde{b}_{L}},m_{h}) - D_{SS}(m_{\tilde{b}_{L}},m_{\pi}) + 2
c_{t}^{2} D_{SS}(m_{\tilde{t}_{1}},m_{\pi})
 + 2 s_{t}^{2}
D_{SS}(m_{\tilde{t}_{2}},m_{\pi})]\nonumber \\
&+&  N_{c} h_{t}^{L}[ D_{SS}(m_{b_{L}},m_{\pi})] \nonumber \\
&+& (g_{s_{1}}^{QD} + 3 g_{s_{2}}^{QD}) N_{c}[c_{t}^{2}D_{SS}(m_{\tilde{t}_{1}},m_{\tilde{b}_{R}})+ s_{t}^{2} D_{SS}(m_{\tilde{t}_{2}},m_{\tilde{b}_{R}}) +
D_{SS}(m_{\tilde{b}_{L}}, m_{\tilde{b}_{R}})] \nonumber \\
& + &(g_{s_{1}}^{UD} + 3 g_{s_{2}}^{UD}) N_{c}[s_{t}^{2}D_{SS}(m_{\tilde{t}_{1}},m_{\tilde{b}_{R}}) + c_{t}^{2} D_{SS}(m_{\tilde{t}_{2}}, m_{\tilde{b}_{R}})].
\label{Vheavyphi}
\end{eqnarray}

\subsubsection{$\chi$-direction}

\begin{eqnarray}
V_{2}^{heavy}(\chi) &=&  -{g_{w_{3}}^{2}\over 8}[ c_{q}^{4} D_{SSV}(m_{\tilde{q}_{1}},m_{\tilde{q}_{1}},0)
+ s_{q}^{4} D_{SSV}(m_{H_{1}},m_{H_{1}},0) + 2 c_{q}^{2} s_{q}^{2} D_{SSV}(m_{\tilde{q}_{1}},m_{H_{2}},0) \nonumber \\
&+&
D_{SSV}(m_{\tilde{t}_{L2}},m_{\tilde{t}_{L2}},0) +
D_{SSV}(m_{\tilde{t}_{L3}},m_{\tilde{t}_{L3}},0)\nonumber \\
& +&
c_{q}^{4} D_{SSV}(m_{\tilde{q}_{2}},m_{\tilde{q}_{2}},0)+
s_{q}^{4} D_{SSV}(m_{H_{2}},m_{H_{2}},0) + 2 c_{q}^{2} s_{q}^{2} D_{SSV}(m_{\tilde{q}_{2}},m_{H_{1}},0) \nonumber \\
&+&
D_{SSV}(m_{\tilde{b}_{L2}},m_{\tilde{b}_{L2}},0) +
D_{SSV}(m_{\tilde{b}_{L3}},m_{\tilde{b}_{L3}},0)\nonumber \\
& +&
4(c_{q}^{4} D_{SSV}(m_{\tilde{q}_{1}},m_{\tilde{q}_{2}},0) + s_{q}^{4} D_{SSV}(m_{H_{1}}, m_{H_{2}},0) + 2c_{q}^{2} s_{q}^{2} D_{SSV}(m_{\tilde{q}_{1}},m_{H_{1}},0)\nonumber \\
&+& 2 c_{q}^{2} s_{q}^{2} D_{SSV}(m_{\tilde{q}_{2}},m_{H_{2}},0)+ D_{SSV}(m_{\tilde{t}_{L2}},m_{\tilde{b}_{L2}},0) +
D_{SSV}(m_{\tilde{t}_{L3}},m_{\tilde{b}_{L3}},0) )]\nonumber \\
&-& g_{s_{3}}^{2} {1\over 4}[ 2(N_{c} -1)(c_{q}^{2} D_{SSV}(m_{\tilde{q}_{1}},m_{\tilde{t}_{L2}},m_{G})
+ s_{q}^{2} D_{SSV}(m_{H_{2}},m_{\tilde{t}_{L2}},m_{G})) \nonumber \\
 &+& {N_{c}-1\over N_{c}} (c_{q}^{4}D_{SSV}(m_{\tilde{q}_{1}},m_{\tilde{q}_{1}},\overline{m}_{G}) + s_{q}^{4} D_{SSV}(m_{H_{2}},m_{H_{2}},\overline{m}_{G})
\nonumber \\
&+& 2 s_{q}^{2} c_{q}^{2} D_{SSV}(m_{\tilde{q}_{1}},m_{H_{2}},\overline{m}_{G}))
+ 
{1\over N_{c}} D_{SSV}(m_{\tilde{t}_{L2}},m_{\tilde{t}_{L2}},\overline{m}_{G})\nonumber \\
&+&
N_{c}(N_{c}-2) D_{SSV}(m_{\tilde{t}_{L2}},m_{\tilde{t}_{L2}},0)\nonumber \\
&+& 2(N_{c} -1) (c_{q}^{2} D_{SSV}(m_{\tilde{q}_{2}},m_{\tilde{b}_{L2}},m_{G})
+ s_{q}^{2} D_{SSV}(m_{H_{1}},m_{\tilde{b}_{L2}},m_{G}))
 \nonumber \\
&+& {N_{c}-1\over N_{c}} (c_{q}^{4}D_{SSV}(m_{\tilde{q}_{2}},m_{\tilde{q}_{2}},\overline{m}_{G}) + s_{q}^{4}D_{SSV}(m_{H_{1}},m_{H_{1}},\overline{m}_{G}) 
+ 2 s_{q}^{2} c_{q}^{2}D_{SSV}(m_{\tilde{q}_{2}},m_{H_{1}},\overline{m}_{G})\nonumber \\
&+& 
{1\over N_{c}} D_{SSV}(m_{\tilde{b}_{L2}},m_{\tilde{b}_{L2}},\overline{m}_{G})+
N_{c}(N_{c}-2) D_{SSV}(m_{\tilde{b}_{L2}},m_{\tilde{b}_{L2}},0)\nonumber \\
&+& 2(N_{c} -1) D_{SSV}(m_{\tilde{b}_{R1}},m_{\tilde{b}_{R2}},m_{G})
 + {N_{c}-1\over N_{c}} D_{SSV}(m_{\tilde{b}_{R1}},m_{\tilde{b}_{R1}},\overline{m}_{G})\nonumber \\
& +& {1\over N_{c}} D_{SSV}(m_{\tilde{b}_{R2}},m_{\tilde{b}_{R2}},\overline{m}_{G})+
N_{c}(N_{c}-2) D_{SSV}(m_{\tilde{b}_{R2}},m_{\tilde{b}_{R2}},0)]\nonumber \\
&-& {1\over 2} (( h_{t}^{QU} + g_{s_{1}}^{QU} + g_{s_{2}}^{QU})s_{q}^{2}\chi + \sqrt 2 \overline{X}_{t} h_{t} c_{q} s_{q})^{2}[D_{SSS}(m_{u},m_{H_{1}},m_{H_{1}}) +
D_{SSS}(m_{u}, m_{H_{2}},m_{H_{2}})] \nonumber \\
&-& {1\over 2}((h_{t}^{QU} + g_{s_{1}}^{QU} + g_{s_{2}}^{QU})c_{q}^{2}\chi - \sqrt 2 \overline{X}_{t} h_{t} c_{q} s_{q})^{2}
[D_{SSS}(m_{u},m_{\tilde{q}_{1}},m_{\tilde{q}_{1}}) \nonumber \\
&+&
D_{SSS}(m_{u},m_{\tilde{q}_{2}},m_{\tilde{q}_{2}})]\nonumber \\
&-&
2[(h_{t}^{QU} + g_{s_{1}}^{QU}){\chi \over \sqrt2}c_{q} - s_{q} \overline{X}_{t} h_{t}]^{2} [D_{SSS}(m_{\omega},m_{\tilde{q}_{1}},m_{\tilde{t}_{L2}}) 
+ D_{SSS}(m_{\omega},m_{\tilde{q}_{2}},m_{\tilde{b}_{L2}}]\nonumber \\
&-&
2[(h_{t}^{QU} + g_{s_{1}}^{QU}){\chi \over \sqrt2}s_{q} + c_{q} \overline{X}_{t} h_{t}]^{2} [D_{SSS}(m_{\omega},m_{H_{2}},m_{\tilde{t}_{L2}})) 
+D_{SSS}(m_{\omega},m_{H_{1}},m_{\tilde{b}_{L2}})]
 \nonumber \\
&-& ( c_{2q}{\overline{X}_{t}\over \sqrt 2} + s_{q} c_{q} (h_{t}^{QU} + g_{s_{1}}^{QU} + g_{s_{2}}^{QU})\chi)^{2} D_{SSS}(m_{u},m_{H_{2}}, m_{q_{1}})
\nonumber \\
&-& ( c_{2q} {\overline{X}_{t}\over \sqrt 2} -  s_{q} c_{q} (h_{t}^{QU} + g_{s_{1}}^{QU} + g_{s_{2}}^{QU})\chi)^{2} D_{SSS}(m_{u},m_{H_{1}}, m_{q_{2}})
\nonumber \\
&-& [c_{2q} {\overline{X}_{t}\over \sqrt 2} h_{t}]^{2}(D_{SSS}(m_{\omega},m_{H_{1}}, m_{q_{2}}) + D_{SSS}(m_{\omega},m_{H_{2}}, m_{q_{1}}))\nonumber \\
&-&{\chi^{2}\over 2}[ 2(g_{s_{2}}^{QU})^{2}D_{SSS}(m_{u},m_{\tilde{t}_{L2}},m_{\tilde{t}_{L2}})
 \nonumber \\
& +& 2(g_{s_{2}}^{QU})^{2}D_{SSS}(m_{u},m_{\tilde{b}_{L2}},m_{\tilde{b}_{L2}}) \nonumber \\
&+&(  g_{s_{1}}^{UD} + g_{s_{2}}^{UD})^{2}
D_{SSS}(m_{u},m_{\tilde{b}_{R1}},m_{\tilde{b}_{R1}}) +
2(g_{s_{1}}^{UD})^{2}D_{SSS}(m_{\omega},m_{\tilde{b}_{R1}},m_{\tilde{b}_{R2}})\nonumber \\
&+& 2(g_{s_{2}}^{UD})^{2}D_{SSS}(m_{u},m_{\tilde{b}_{R2}},m_{\tilde{b}_{R2}})] \nonumber \\
&-& {g_{s_{3}}^{2}\over 8}[ 2N_{c}(N_{c}-2) D_{SV}(m_{\tilde{t}_{L2}},0) +
(N_{c} -1)[2D_{SV}(m_{\tilde{t}_{L2}},m_{G}) +2 (c_{q}^{2}D_{SV}(m_{\tilde{q}_{1}},m_{G}) \nonumber \\
&+& s_{q}^{2} D_{SV}(m_{H_{2}},m_{G}))]
+ {1\over N_{c}}[2D_{SV}(m_{\tilde{t}_{L2}},\overline{m}_{G})
+2(c_{q}^{2}D_{SV}(m_{\tilde{q}_{1}},\overline{m}_{G}) + s_{q}^{2} D_{SV}(m_{H_{2}},\overline{m}_{G}))] \nonumber \\
&+& 2N_{c}(N_{c}-2) D_{SV}(m_{\tilde{b}_{L2}},0) +
(N_{c} -1)[2D_{SV}(m_{\tilde{b}_{L2}},m_{G}) +2 (c_{q}^{2}D_{SV}(m_{\tilde{q}_{2}},m_{G}) \nonumber \\
&+& s_{q}^{2} D_{SV}(m_{H_{1}},m_{G}))] 
+ {1\over N_{c}}[2D_{SV}(m_{\tilde{b}_{L2}},\overline{m}_{G})
+2(c_{q}^{2}D_{SV}(m_{\tilde{q}_{2}},\overline{m}_{G}) + s_{q}^{2} D_{SV}(m_{H_{1}},\overline{m}_{G}))] \nonumber \\
&+& 2N_{c}(N_{c}-2) D_{SV}(m_{\tilde{b}_{R2}},0) +
(N_{c} -1)[2D_{SV}(m_{\tilde{b}_{R2}},m_{G}) +2 D_{SV}(m_{\tilde{b}_{R1}},m_{G})] \nonumber \\
&+& {1\over N_{c}}[2D_{SV}(m_{\tilde{b}_{R2}},\overline{m}_{G})
+2D_{SV}(m_{\tilde{b}_{R1}},\overline{m}_{G})] \nonumber \\
&-& {3\over 8} g_{w_{3}}^{2}[c_{q}^{2}D_{SV}(m_{\tilde{q}_{1}},0) + D_{SV}(m_{\tilde{t}_{L2}},0) + D_{SV}(m_{\tilde{t}_{L3}},0) \nonumber \\
&+& c_{q}^{2} D_{SV}(m_{\tilde{q}_{2}},0) + D_{SV}(m_{\tilde{b}_{L2}},0) + D_{SV}(m_{\tilde{b}_{L3}},0) + s_{q}^{2} D_{SV}(m_{H_{1}},0) + s_{q}^{2} D_{SV}(m_{H_{2}},0)]\nonumber \\
&+& \biggl(\Lambda_{1} + \lambda_{Q_{3}}\biggr)[2 c_{q}^{4} D_{SS}(m_{\tilde{q}_{1}},m_{\tilde{q}_{1}}) +
2 s_{q}^{4} D_{SS}(m_{h_{2}},m_{h_{2}}) +
6 s_{q}^{2} c_{q}^{2} D_{SS}(m_{h_{2}},m_{\tilde{q}_{1}}) \nonumber \\
&+& 4(s_{q}^{2} D_{SS}(m_{h_{2}},m_{\tilde{t}_{L2}}) + 
c_{q}^{2} D_{SS}(m_{\tilde{q}_{1}},m_{\tilde{t}_{L2}}) +
6 D_{SS}(m_{\tilde{t}_{L2}},m_{\tilde{t}_{L2}}) \nonumber \\
&+&2 c_{q}^{4} D_{SS}(m_{\tilde{q}_{2}},m_{\tilde{q}_{2}}) +
2 s_{q}^{4} D_{SS}(m_{h_{1}},m_{h_{1}}) +
6 s_{q}^{2} c_{q}^{2} D_{SS}(m_{h_{1}},m_{\tilde{q}_{2}}) \nonumber \\
&+& 4(s_{q}^{2} D_{SS}(m_{h_{1}},m_{\tilde{t}_{L2}}) + 
c_{q}^{2} D_{SS}(m_{\tilde{q}_{2}},m_{\tilde{t}_{L2}}) +
6 D_{SS}(m_{\tilde{b}_{L2}},m_{\tilde{b}_{L2}})] \nonumber \\
&+& \lambda_{D_{3}}[8 D_{SS}(m_{\tilde{b}_{R1}},m_{\tilde{b}_{R1}}) +
24 D_{SS}(m_{\tilde{b}_{R2}},m_{\tilde{b}_{R2}}) + 16 D_{SS}(m_{\tilde{b}_{R1}},m_{\tilde{b}_{R2}})]\nonumber \\
&+& {1\over 2} (h_{t}^{QU}+ g_{s_{1}}^{QU})[c_{q}^{2} D_{SS}(m_{\tilde{q}_{1}},m_{u}) + c_{q}^{2} D_{SS}(m_{\tilde{q}_{1}},m_{\omega})
+ s_{q}^{2} D_{SS}(m_{h_{2}},m_{u}) + s_{q}^{2} D_{SS}(m_{h_{2}},m_{\omega})\nonumber \\
& +& 2 D_{SS}(m_{\tilde{t}_{L2}},m_{\omega})
+ 2 D_{SS}(m_{\tilde{t}_{L3}},m_{\omega})] \nonumber \\ 
&+& {1\over 2} g_{s_{2}}^{QU}[c_{q}^{2}D_{SS}(m_{u},m_{\tilde{q}_{1}})
+ 5 c_{q}^{2} D_{SS}(m_{\omega},m_{\tilde{q}_{1}}) +
s_{q}^{2}D_{SS}(m_{u},m_{h_{2}})\nonumber \\
&+& 5 s_{q}^{2} D_{SS}(m_{\omega},m_{h_{2}}) +
D_{SS}(m_{u},m_{\tilde{t}_{L2}}) \nonumber \\
&+& 5 D_{SS}(m_{\omega},m_{\tilde{t}_{L2}}) +
D_{SS}(m_{u},m_{\tilde{t}_{L3}})
+ 5 D_{SS}(m_{\omega},m_{\tilde{t}_{L3}})]\nonumber \\
&+& {1\over 2} (h_{t}^{QU} + g_{s_{1}}^{QU})[c_{q}^{2} D_{SS}(m_{\tilde{q}_{2}},m_{u}) + c_{q}^{2} D_{SS}(m_{\tilde{q}_{2}},m_{\omega})\nonumber \\
&+& s_{q}^{2} D_{SS}(m_{h_{1}},m_{u}) + s_{q}^{2} D_{SS}(m_{h_{1}},m_{\omega}) + 2 D_{SS}(m_{\tilde{t}_{L2}},m_{\omega})
+ 2 D_{SS}(m_{\tilde{t}_{L3}},m_{\omega})] \nonumber \\ 
&+& {1\over 2} g_{s_{2}}^{QU}[c_{q}^{2}D_{SS}(m_{u},m_{\tilde{q}_{2}})
+ 5 c_{q}^{2} D_{SS}(m_{\omega},m_{\tilde{q}_{2}}) +
s_{q}^{2}D_{SS}(m_{u},m_{h_{1}})\nonumber \\
&+& 5 s_{q}^{2} D_{SS}(m_{\omega},m_{h_{1}}) +
D_{SS}(m_{u},m_{\tilde{b}_{L2}}) \nonumber \\
&+& 5 D_{SS}(m_{\omega},m_{\tilde{b}_{L2}}) +
D_{SS}(m_{u},m_{\tilde{b}_{L3}})
+ 5 D_{SS}(m_{\omega},m_{\tilde{b}_{L3}})]\nonumber \\
&+& \biggl( h_{t}^{L} + \Lambda_{3} + \Lambda_{4}^{s}\biggr) \biggl[2 c_{q}^{2} s_{q}^{2}[D_{SS}(m_{h_{1}},m_{h_{1}}) \nonumber \\
&+& D_{SS}(m_{h_{2}},m_{h_{2}})
+  D_{SS}(m_{h_{1}},m_{h_{2}}) +  D_{SS}(m_{\tilde{q}_{1}},m_{\tilde{q}_{2}})
+ D_{SS}(m_{\tilde{q}_{1}},m_{\tilde{q}_{1}})\nonumber \\
& +&
D_{SS}(m_{\tilde{q}_{2}},m_{\tilde{q}_{2}})]
+ c_{q}^{2}[D_{SS}(m_{\tilde{t}_{L2}},m_{h_{2}})  + D_{SS}(m_{\tilde{t}_{L3}},m_{h_{2}}) \nonumber \\
&+&  D_{SS}(m_{\tilde{b}_{L2}},m_{h_{1}})  + D_{SS}(m_{\tilde{b}_{L3}},m_{h_{1}})] + s_{q}^{2} [D_{SS}(m_{\tilde{t}_{L2}},m_{\tilde{q}_{1}})  + D_{SS}(m_{\tilde{t}_{L3}},m_{\tilde{q}_{1}}) \nonumber \\
&+&  D_{SS}(m_{\tilde{b}_{L2}},m_{\tilde{q}_{2}})  + D_{SS}(m_{\tilde{b}_{L3}},m_{\tilde{q}_{2}})]+
 c_{2q}^{2}[D_{SS}(m_{\tilde{q}_{1}},m_{h_{2}})
+ D_{SS}(m_{\tilde{q}_{2}},m_{h_{1}})] \biggr]\nonumber \\
&+& \biggl( (\Lambda_{3} + \Lambda_{4}^{c})(c_{q}^{4} + s_{q}^{4})  - 2 s_{q}^{2} c_{q}^{2}(\Lambda_{4}^{s} + h_{t}^{L} )\biggr)[D_{SS}(m_{\tilde{q}_{1}},m_{h_{1}}) + D_{SS}(m_{\tilde{q}_{2}},m_{h_{2}})]\nonumber \\
&+& + 2 \Lambda_{4}^{c} s_{q}^{2} c_{q}^{2} [D{SS}(m_{h_{1}}, m_{h_{2}}) +
D_{SS}(m_{\tilde{q}_{1}}, m_{\tilde{q}_{2}})] \nonumber \\
&- &  \biggl(\Lambda_{3} + \Lambda_{4}^{c}\biggr)[ c_{q}^{2}(D_{SS}(m_{\tilde{b}_{L2}},m_{h_{2}})  + D_{SS}(m_{\tilde{b}_{L3}},m_{h_{2}}) \nonumber \\
&+&  D_{SS}(m_{\tilde{t}_{L2}},m_{h_{1}})  + D_{SS}(m_{\tilde{t}_{L3}},m_{h_{1}})) + s_{q}^{2}(D_{SS}(m_{\tilde{b}_{L2}},m_{\tilde{q}_{1}})  + D_{SS}(m_{\tilde{b}_{L3}},m_{\tilde{q}_{1}}) \nonumber \\
&+&  D_{SS}(m_{\tilde{t}_{L2}},m_{\tilde{q}_{2}})  + D_{SS}(m_{\tilde{t}_{L3}},m_{\tilde{q}_{2}})) ]\nonumber \\
&+& 2 \Lambda_{1} (2-N_{c})[c_{q}^{4} D_{SS}(m_{\tilde{q}_{1}},m_{\tilde{q}_{2}})\nonumber \\
& +& s_{q}^{4} D_{SS}(m_{h_{1}},m_{h_{2}}) + s_{q}^{2} c_{q}^{2} D_{SS}(m_{h_{2}},m_{\tilde{q}_{2}}) + s_{q}^{2} c_{q}^{2} D_{SS}(m_{h_{1}}, m_{\tilde{q}_{1}}) \nonumber \\
&+&  D_{SS}(m_{\tilde{t}_{L2}},m_{\tilde{b}_{L2}}) + D_{SS}(m_{\tilde{t}_{L3}},m_{\tilde{b}_{L3}})] \nonumber \\
&+&  g_{s_{1}}^{QQ}[ c_{q}^{4}D_{SS}(m_{\tilde{q}_{1}},m_{\tilde{q}_{2}}) +
s_{q}^{4} D_{SS}(m_{h_{1}}, m_{h_{2}}) + s_{q}^{2} c_{q}^{2} (D_{SS}(m_{\tilde{q}_{1}}, m_{h_{1}})\nonumber \\
& +& D_{SS}(m_{\tilde{q}_{2}}, m_{h_{2}})) +
D_{SS}(m_{\tilde{t}_{L2}}, m_{\tilde{b}_{L2}})
+ D_{SS}(m_{\tilde{b}_{L3}},m_{\tilde{t}_{L3}})]\nonumber \\
&+& g_{s_{2}}^{QQ}[ c_{q}^{4}D_{SS}(m_{\tilde{q}_{1}},m_{\tilde{q}_{2}}) +
s_{q}^{4} D_{SS}(m_{h_{1}}, m_{h_{2}}) + s_{q}^{2} c_{q}^{2} (D_{SS}(m_{\tilde{q}_{1}}, m_{h_{1}}) \nonumber \\
&+& D_{SS}(m_{\tilde{q}_{2}}, m_{h_{2}}))+
c_{q}^{2}D_{SS}(m_{\tilde{q}_{1}},m_{\tilde{b}_{L2}})+ s_{q}^{2} D_{SS}(m_{h_{2}}, m_{\tilde{b}_{L2}})\nonumber \\
& +& c_{q}^{2} D_{SS}(m_{\tilde{q}_{1}},m_{\tilde{b}_{L3}}) + s_{q}^{2} D_{SS}(m_{h_{2}}, m_{\tilde{b}_{L3}})\nonumber \\
&+& c_{q}^{2} D_{SS}(m_{\tilde{t}_{L2}},m_{\tilde{q}_{2}}) + s_{q}^{2} D_{SS}(m_{\tilde{t}_{L2}}, m_{h_{1}}) + D_{SS}(m_{\tilde{t}_{L2}},m_{\tilde{b}_{L2}})+
D_{SS}(m_{\tilde{t}_{L2}},m_{\tilde{b}_{L3}})\nonumber \\
& +& c_{q}^{2}D_{SS}(m_{\tilde{t}_{L3}},m_{\tilde{q}_{2}}) + s_{q}^{2} D_{SS}(m_{\tilde{t}_{L3}}, m_{h_{1}})
+ D_{SS}(m_{\tilde{t}_{L3}},m_{\tilde{b}_{L2}})\nonumber \\
&+& D_{SS}(m_{\tilde{t}_{L3}},m_{\tilde{b}_{L3}})\nonumber \\
&+&  g_{s_{1}}^{QD}[c_{q}^{2} D_{SS}(m_{\tilde{q}_{1}},m_{\tilde{b}_{R1}})
+ s_{q}^{2} D_{SS}(m_{h_{2}}, m_{\tilde{b}_{R1}}) + D_{SS}(m_{\tilde{t}_{L2}},m_{\tilde{b}_{R2}})\nonumber \\
&+& D_{SS}(m_{\tilde{t}_{L3}},m_{\tilde{b}_{R3}}) + c_{q}^{2} D_{SS}(m_{\tilde{q}_{2}},m_{\tilde{b}_{R1}}) + s_{q}^{2} D_{SS}(m_{h_{1}}, m_{\tilde{b}_{R1}}) \nonumber \\
& +& D_{SS}(m_{\tilde{b}_{L2}},m_{\tilde{b}_{R2}})
+ D_{SS}(m_{\tilde{b}_{L3}},m_{\tilde{b}_{R3}})]\nonumber \\
&+& g_{s_{2}}^{QD}(c_{q}^{2} D_{SS}(m_{\tilde{q}_{1}},m_{\tilde{b}_{R1}})
+ s_{q}^{2} D_{SS}(m_{h_{2}}, m_{\tilde{b}_{R1}}) \nonumber \\
&+&c_{q}^{2} D_{SS}(m_{\tilde{q}_{1}},m_{\tilde{b}_{R2}})
+ s_{q}^{2} D_{SS}(m_{h_{2}}, m_{\tilde{b}_{R2}})  + c_{q}^{2} D_{SS}(m_{\tilde{q}_{1}},m_{\tilde{b}_{R3}})
+ s_{q}^{2} D_{SS}(m_{h_{2}}, m_{\tilde{b}_{R3}}) \nonumber \\ 
&+& D_{SS}(m_{\tilde{t}_{L2}},m_{\tilde{b}_{R1}}) + D_{SS}(m_{\tilde{t}_{L2}},m_{\tilde{b}_{R2}})+
D_{SS}(m_{\tilde{t}_{L2}},m_{\tilde{b}_{R3}})\nonumber \\
& +& D_{SS}(m_{\tilde{t}_{L3}},m_{\tilde{b}_{R1}})
+ D_{SS}(m_{\tilde{t}_{L3}},m_{\tilde{b}_{R2}})\nonumber \\
&+& D_{SS}(m_{\tilde{t}_{L3}},m_{\tilde{b}_{R3}}) 
\nonumber \\
&+&c_{q}^{2} D_{SS}(m_{\tilde{q}_{2}},m_{\tilde{b}_{R1}})
+ s_{q}^{2} D_{SS}(m_{h_{1}}, m_{\tilde{b}_{R1}}) \nonumber \\
&+&c_{q}^{2} D_{SS}(m_{\tilde{q}_{2}},m_{\tilde{b}_{R2}})
+ s_{q}^{2} D_{SS}(m_{h_{1}}, m_{\tilde{b}_{R2}})  + c_{q}^{2} D_{SS}(m_{\tilde{q}_{2}},m_{\tilde{b}_{R3}})
+ s_{q}^{2} D_{SS}(m_{h_{1}}, m_{\tilde{b}_{R3}}) \nonumber \\
&+& D_{SS}(m_{\tilde{b}_{L2}},m_{\tilde{b}_{R1}}) + D_{SS}(m_{\tilde{b}_{L2}},m_{\tilde{b}_{R2}})+
D_{SS}(m_{\tilde{b}_{L2}},m_{\tilde{b}_{R3}}) + D_{SS}(m_{\tilde{b}_{L3}},m_{\tilde{b}_{R1}})\nonumber \\
&+& D_{SS}(m_{\tilde{b}_{L3}},m_{\tilde{b}_{R2}})
+ D_{SS}(m_{\tilde{b}_{L3}},m_{\tilde{b}_{R3}})]\nonumber \\
&+& {1\over 2} g_{s_{1}}^{UD}[D_{SS}(m_{u},m_{\tilde{b}_{R1}}) + D_{SS}(m_{\omega},m_{\tilde{b}_{R1}}) + 2 D_{SS}(m_{\omega},m_{\tilde{b}_{R2}})
+ 2 D_{SS}(m_{\omega},m_{\tilde{b}_{R3}})]\nonumber \\
&+& {1\over 2} g_{s_{2}}^{UD}[D_{SS}(m_{u},m_{\tilde{b}_{R1}})+
D_{SS}(m_{u},m_{\tilde{b}_{R2}}) + D_{SS}(m_{u},m_{\tilde{b}_{R3}})\nonumber \\
&+& 5 D_{SS}(m_{\omega},m_{\tilde{b}_{R1}}) +5 D_{SS}(m_{\omega},m_{\tilde{b}_{R2}})+
5D_{SS}(m_{\omega},m_{\tilde{b}_{R3}})].
\label{V2heavy}
\end{eqnarray}

\section{Zero-temperature renormalization}

The most important zero-temperature renormalization  effects with
respect to our calculation 
concern the mass parameters.
We will not go into the details of the renormalization,
but refer the reader to the literature in which
the pole masses for the relevant particles of our calculation
have been obtained considering the full particle spectrum of the
MSSM \cite{bagger, donini}. We use the
expressions given in ref. \cite{bagger}, keeping only the
top Yukawa coupling, in the appropriate (large-$m_{A}$) limit. In this limit
the relevant expression for the one-loop corrected Higgs mass
is given by \cite{Ellis, Carena3, Hempfling}:

\begin{eqnarray}
m_{h}^2 &=& m_{Z}^2 \cos^2 2\beta   + { 3 g_{w}^2 m_{t}^4\over16 \pi^2 m_{W}^2}[(2\ln{m_{\tilde{t}_{1}}m_{\tilde{t}_{2}}\over m_{t}^{2}}
+ ({m_{\tilde{t}_{2}}^2- m_{\tilde{t}_{1}}^2 \over 4 m_{t}^2 } \sin^{2} 2 \alpha_{t})^2 f(m_{\tilde{t}_{1}}^2,m_{\tilde{t}_{2}}^2) \nonumber \\
& +&  2(m_{\tilde{t}_{2}}^2- m_{\tilde{t}_{1}}^2)/(2 m_{t}^2) \
\sin^{2}2 \alpha_{t} \ln[m_{\tilde{t}_{2}}/m_{\tilde{t}_{1}}])]
\label{Higgsmass}
\end{eqnarray}
where $f(x,y) = 2 - {x+y\over (x-y)}\ln{x\over y}$.

 In principle, there exists a metastable region  in which the
colour-breaking minimum is lower than the physical one at zero temperature.
 The constraint for absolute stability
 can be obtained by studying the effective potential
at zero temperature \cite{Carena1, Carena2}. This gives the constraint
$-m_{U}^{2} \leq  (m_{U}^{c})^{2}$, where

\begin{equation}
m_{U}^{c} = \biggl( {m_{h}^{2} v^{2} g_{s}^{2}\over 12}\biggr)^{1/4}.
\label{zeroTconst}
\end{equation}

\vspace{.3in}
\noindent
{\bf Acknowledgements}

\noindent
I would like to thank S. Davidson, J.R. Espinosa, M. Laine  and C. Wagner
 for many useful discussions.

\begin{figure}[ht]
\vskip -10 pt
\epsfxsize=6in
\epsfysize=8in
\epsffile{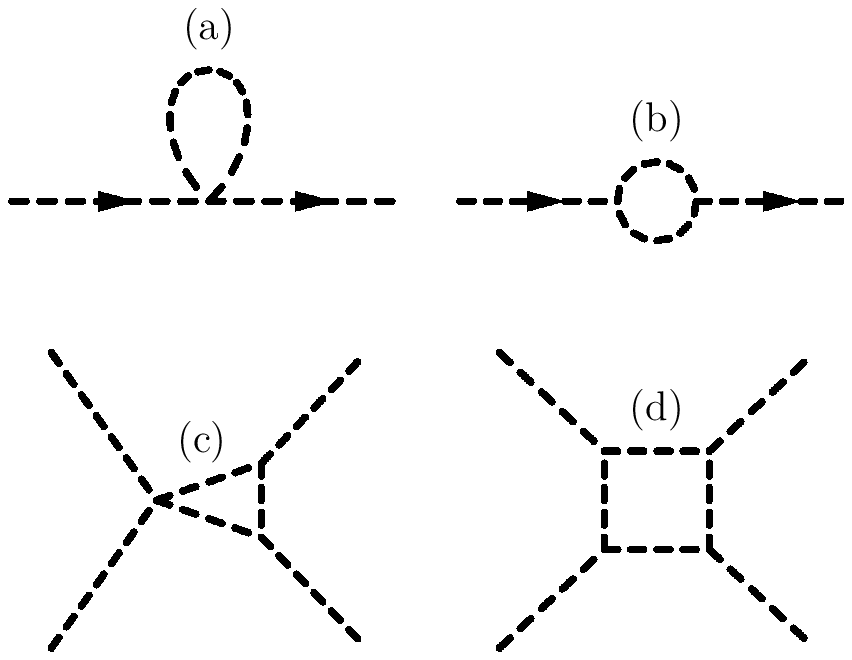}
\vskip -100 pt
\caption{Feynman diagrams contributing to the two- and four-point Green functions. }
\label{Tcmt1}
\end{figure}

\begin{figure}[ht]
\vskip -10 pt
\epsfxsize=4in
\epsfysize=6in
\epsffile{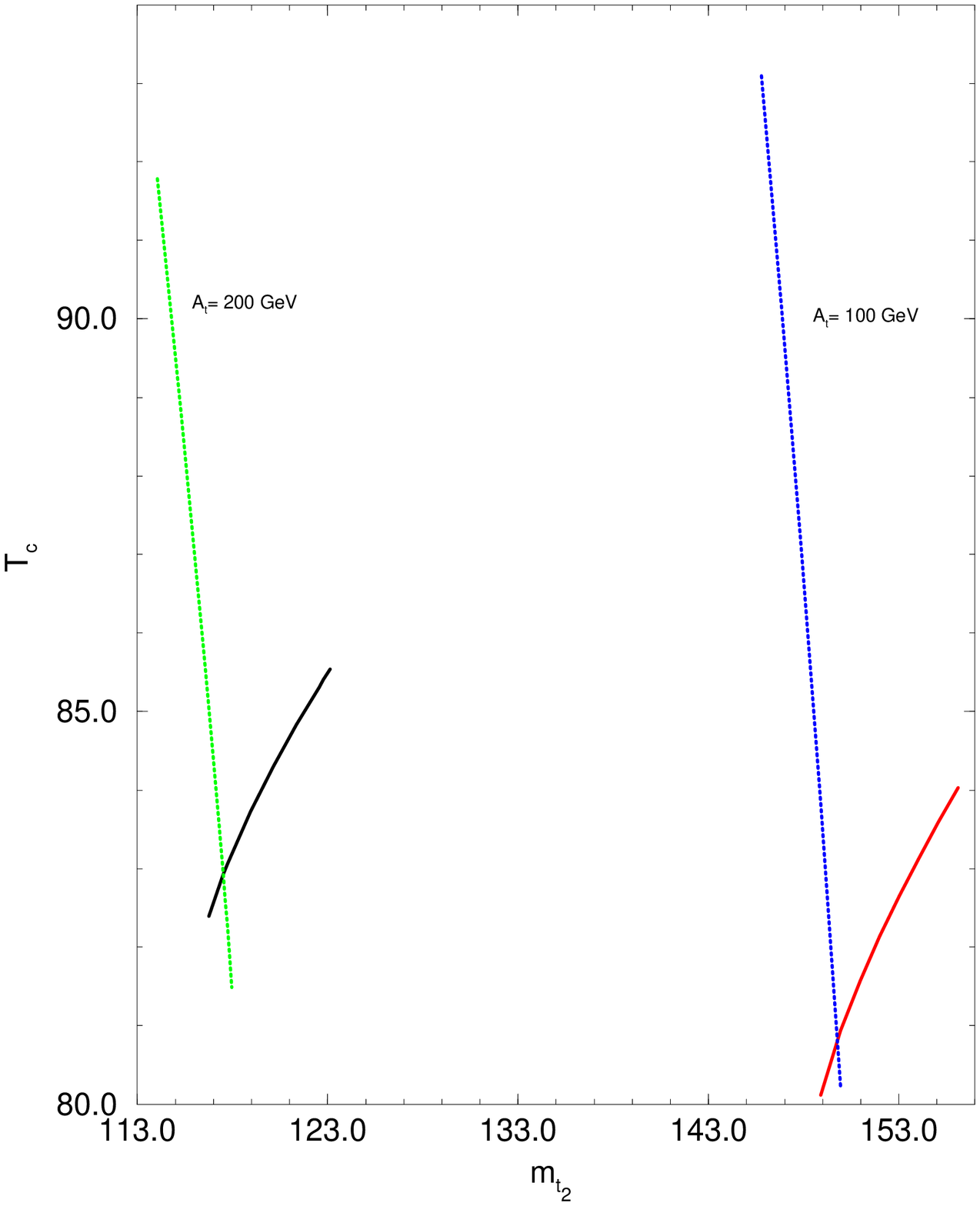}
\vskip -10 pt
\caption{Critical temperatures in the $\phi$- (solid) and $\chi$- (dotted) directions
as functions of $m_{\tilde{t}_{2}}$ for $\tan\beta =5$ and  $m_{Q} = 300$ GeV. }
\label{Tcmt1}
\end{figure}

\begin{figure}
\vskip -10 pt
\epsfxsize=4in
\epsfysize=6in
\epsffile{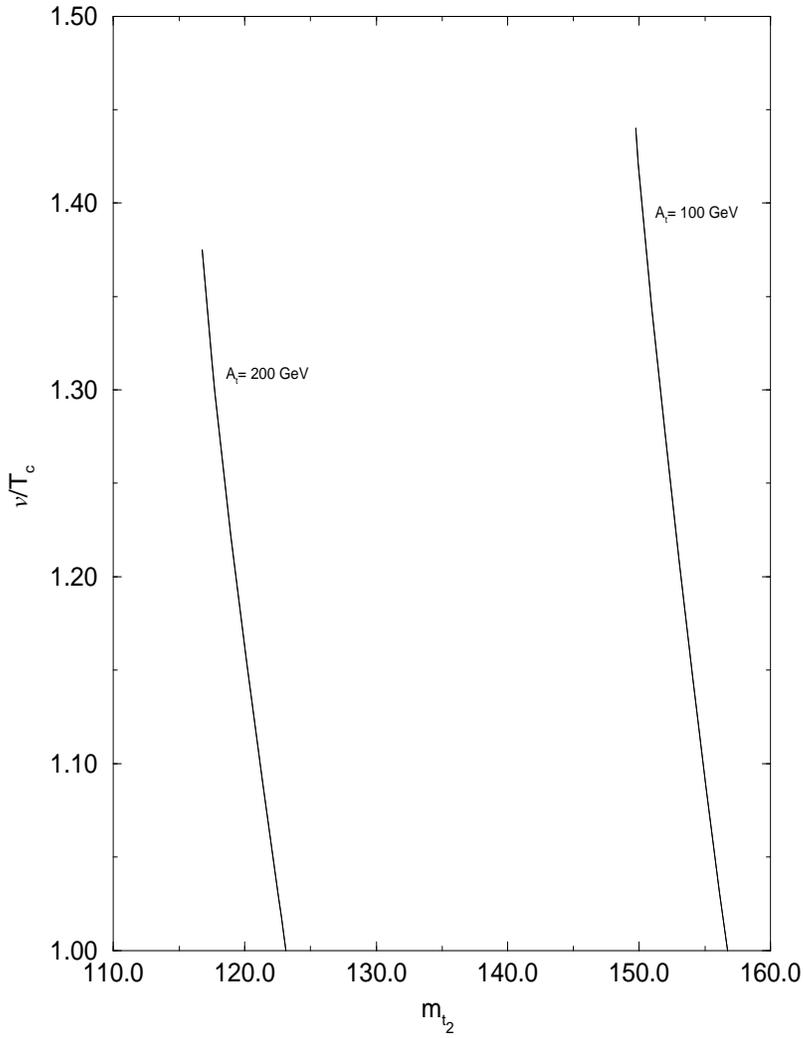}
\vskip -10 pt
\caption{Plot of ${v\over T}$ as a function of $m_{\tilde{t}_{1}}$ in the $\phi$-  direction for $\tan\beta =5$, $m_{Q} = 300$ GeV and $A_{t} =100, 200$ GeV. }
\label{vTmt1}
\end{figure}

\begin{figure}
\vskip -180 pt
\epsfxsize=4in
\epsfysize=6in
\epsffile{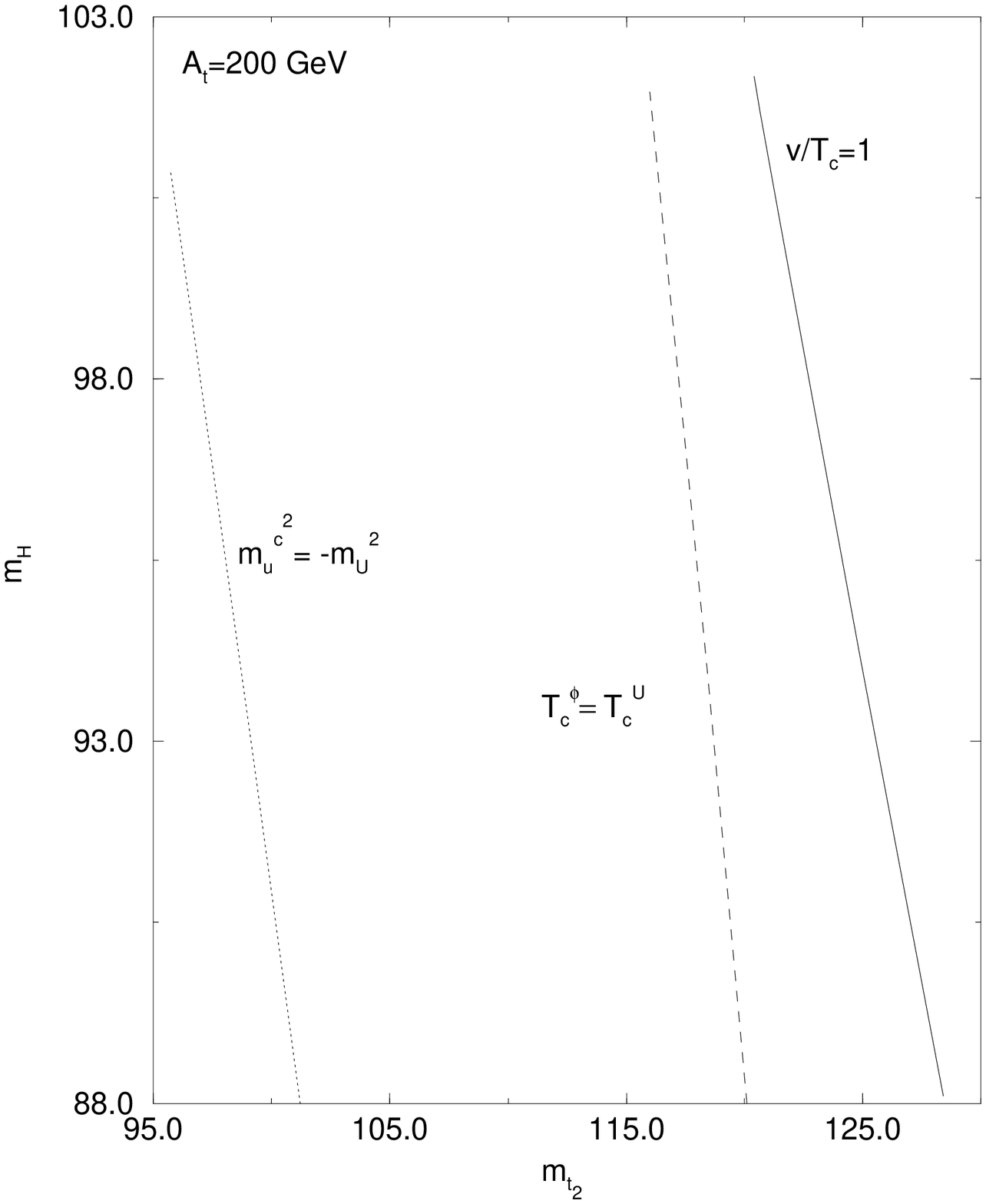}
\vskip -10 pt
\caption{Allowed region in $m_{h}$-$m_{\tilde{t}_{2}}$ plane for $m_{Q}=300$ GeV and ${X_{t}\over \sin\beta} = 200$ GeV. To the left of the solid line
there is a sufficiently strong first-order phase transition, to the right of the dotted line
the physical vacuum is absolutely stable. The dashed line separates the region for which
a two-stage phase transition can occur. }
\label{comp1}
\end{figure}

\begin{figure}
\vskip -180 pt
\epsfxsize=4in
\epsfysize=6in
\epsffile{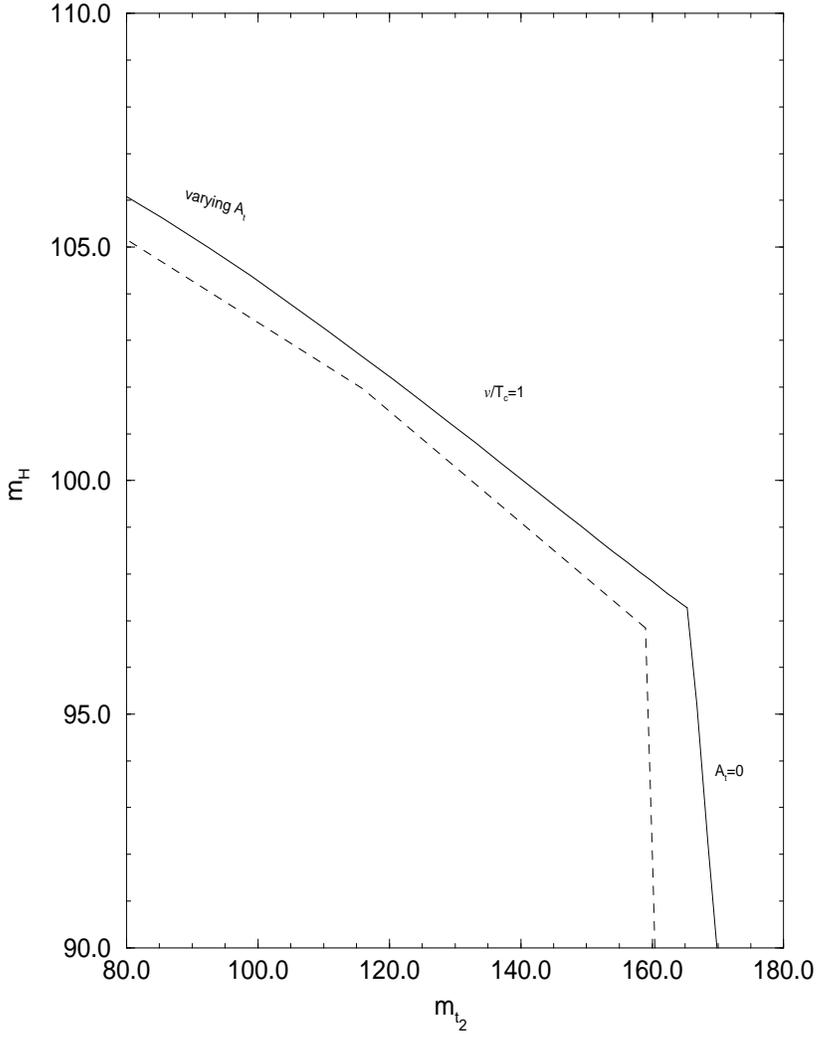}
\vskip -10 pt
\caption{Allowed region of parameter space in the $m_{h}$-$m_{\tilde{t}_{2}}$ plane for $m_{Q}=300$ GeV, varying
$\tilde{A}_{t}$ and $\tan\beta$. The dashed line is defined when the critical temperatures in
the $\phi$- and $\chi$- directions are equal for the same variations of $\tan\beta$ and
$\tilde{A}_{t}$.
 }
\label{vteffec}
\end{figure}

\begin{figure}
\vskip -180 pt
\epsfxsize=4in
\epsfysize=6in
\epsffile{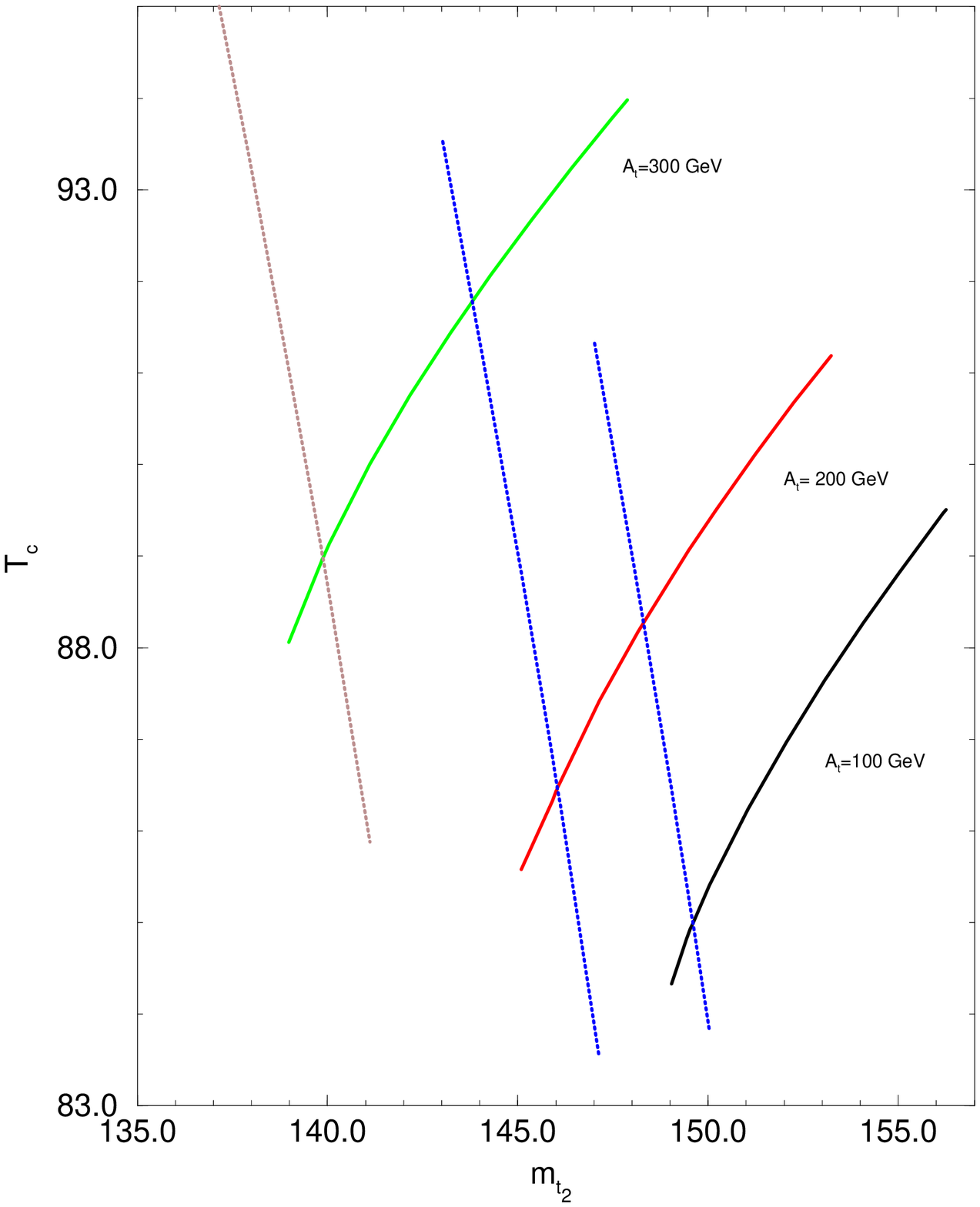}
\vskip -10 pt
\caption{Critical temperatures in the $\phi$- (solid) and $\chi$- (dotted) directions
as functions of $m_{\tilde{t}_{2}}$ for $\tan\beta =5$,  $m_{Q} = 1$ TeV and $\tilde{A}_{t} = 100, 200, 300$ GeV. }
\label{TcmQlarge}
\end{figure}

\begin{figure}
\vskip -180 pt
\epsfxsize=4in
\epsfysize=6in
\epsffile{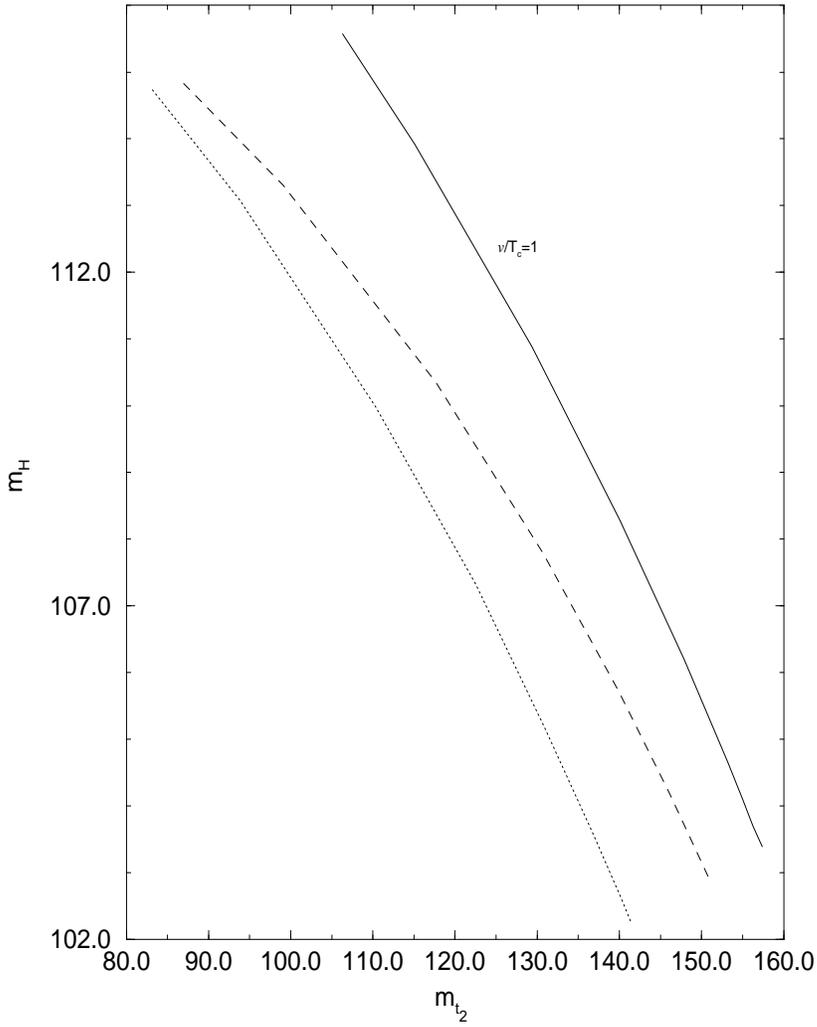}
\vskip -10 pt
\caption{Allowed region of parameter space in the $m_{h}$-$m_{\tilde{t}_{2}}$ plane for $m_{Q}=1$ TeV, 
$0 \leq \tilde{A}_{t} \leq 650$ GeV and $\tan\beta = 5 $. To the left of the solid line
there
 is a sufficiently strong first-order phase transition, to the right of the dotted line
the physical vacuum is absolutely stable. The dashed line separates the region for which
a two-stage phase transition can occur.} 
\label{comp2}
\end{figure}

\begin{figure}
\vskip -180 pt
\epsfxsize=4in
\epsfysize=6in
\epsffile{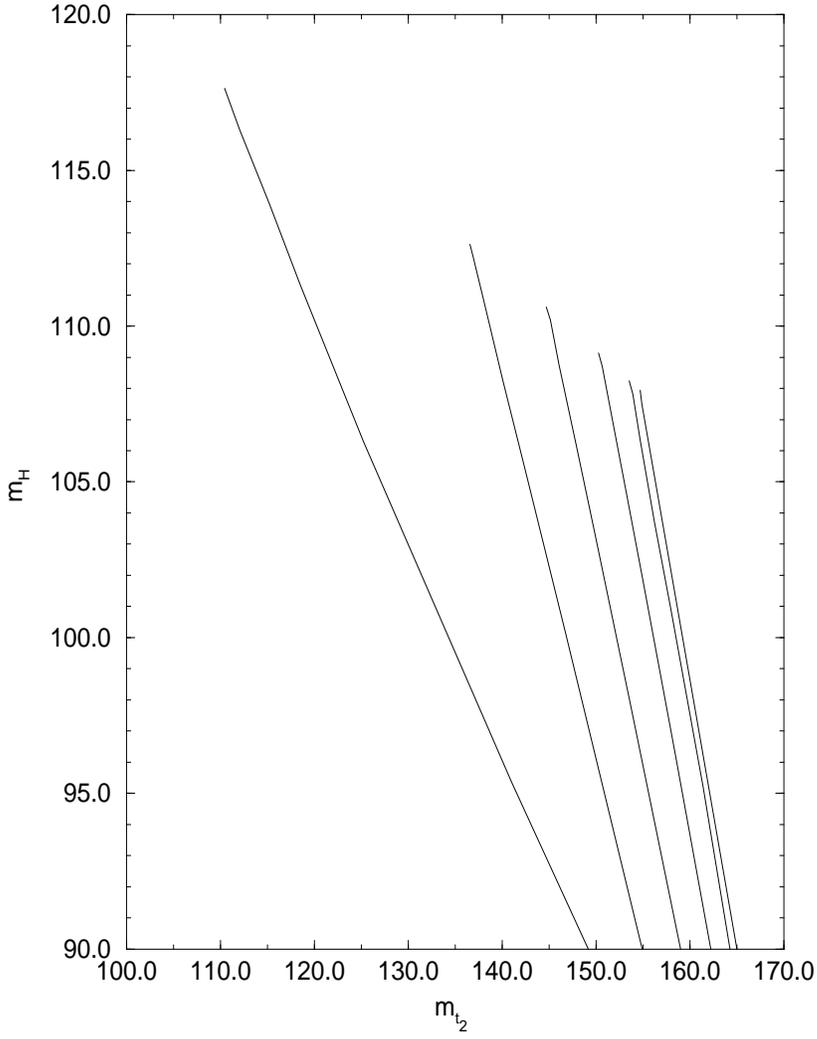}
\vskip -10 pt
\caption{Contours of ${\phi\over T_{c}} =1$ in the $m_{h}$-$m_{\tilde{t}_{2}}$ plane for $m_{Q}=1$ TeV, for
$\tilde{A}_{t} = 0, 100, 200, 300, 400, 600$ GeV. }
\label{param2}
\end{figure}


\end{document}